\documentclass[prb,aps,superscriptaddress,twocolumn,showpacs,dvips]{revtex4}
\usepackage{feynmp}
\usepackage{amssymb}
\usepackage{amsmath}
\usepackage{epsfig}
\usepackage{array,graphicx,subfigure}
\usepackage{mathrsfs}
\usepackage{hyperref}
\usepackage{cases}
\usepackage{bbm}
\usepackage{gensymb}
\usepackage{float}
\usepackage{color}
\usepackage{multirow}
\usepackage{slashbox}
\begin{document}

\title{Interplay of Coulomb interaction and disorder in a two-dimensional semi-Dirac fermion system}

\author{Peng-Lu Zhao}
\affiliation{Department of Modern Physics, University of Science and
Technology of China, Hefei, Anhui 230026, P. R. China}
\author{Jing-Rong Wang}
\affiliation{High Magnetic Field Laboratory, Hefei Institutes of
Physical Science, Chinese Academy of Sciences, Hefei 230031, P. R.
China}
\author{An-Min Wang}
\affiliation{Department of Modern Physics, University of Science and
Technology of China, Hefei, Anhui 230026, P. R. China}
\author{Guo-Zhu Liu}
\altaffiliation{gzliu@ustc.edu.cn} \affiliation{Department of Modern
Physics, University of Science and Technology of China, Hefei, Anhui
230026, P. R. China}

\begin{abstract}
It was recently found that Coulomb interaction can induce a series
of nontrivial spectral and transport properties in a two-dimensional
semi-Dirac fermion system. Different from graphehe that is basically
an ordinary Fermi liquid, the Coulomb interaction in this system
invalidates the Fermi liquid description over a wide range of energy
scales. We present a systematic renormalization group analysis of
the interplay of Coulomb interaction and quenched disorder, and show
that they have substantial mutual effects on each other, which leads
to a variety of quantum phase transitions and non-Fermi liquid
behaviors. The low-energy behaviors of the system depend sensitively
on the effective strength of Coulomb interaction and disorder.
\end{abstract}

\maketitle


\section{Introduction}\label{Sec_intro}

The physical effects induced by Coulomb interaction and by disorder
are two important topics of theoretical condensed matter physics
\cite{Abrikosov, Giuliani, Lee85, Altshuler, Belitz94, Abrahams01,
Altland02, Castroneto09, Sarma11, Kotov12}. They govern most of the
spectral, thermodynamic, and transport properties of quantum many
body systems, and hence have been extensively studied for more than
half a century \cite{Abrikosov, Giuliani, Lee85, Altshuler,
Belitz94, Abrahams01, Altland02, Castroneto09, Sarma11, Kotov12}.

In the normal metal possessing a finite Fermi surface, the
long-range Coulomb interaction is screened by the particle-hole
excitations and only leads to weak damping of the fermionic
quasiparticles, which renders the validity of Landau's Fermi liquid
(FL) theory \cite{Abrikosov, Giuliani}. The role of Coulomb
interaction is more important in correlated electron systems that
feature isolated Fermi points, such as Dirac and Weyl semimetals
\cite{Castroneto09, Sarma11, Kotov12, Wehling14, Vafek14}. In these
semimetals, the density of states (DOS) of fermion vanishes at the
band touching points, so the Coulomb interaction remains
long-ranged, albeit being dynamically screened \cite{Castroneto09,
Sarma11, Kotov12, Wehling14, Vafek14}. The poorly screened Coulomb
interaction may induce strong Landau damping and invalidate the
ordinary FL theory. This line of thought has stimulated intense
investigations in the past years, mainly in the context of graphene,
a two-dimensional (2D) Dirac semimetal at zero doping. Systematic
renormalization group (RG) analysis showed that Coulomb interaction
does cause strong damping of Dirac fermions, but the quasiparticle
residue $Z_f$ flows to a finite constant in the low-energy regime
\cite{Kotov12, Gonzalez99, Hofmann14}. Therefore, although the
Coulomb interaction is long-ranged, the ground state of 2D Dirac
fermion system is still a normal FL. The robustness of the FL
description relies on the fact that the Coulomb interaction is
turned into a marginally irrelevant perturbation by the
substantially enhanced fermion velocity.

Recently, the effects of Coulomb interaction has been studied in a
number of different kinds of semimetals. The Coulomb interaction is
found to be marginally irrelevant in three-dimensional (3D)
Dirac/Weyl semimetals \cite{Goswami2011PRL, Hosur12,
Throckmorton15}. RG analysis performed up to two-loop order argued
that the renormalized fermion velocity may exhibit non-monotonic
energy dependence in some 3D Dirac/Weyl semimetals
\cite{Throckmorton15}. This is quite different from 2D Dirac
semimetals, where the renormalized velocity grows monotonously as
the energy is lowered. One possible reason for such a difference is
that the electric charge is renormalized in 3D Dirac/Weyl
semimetals, but unrenormalized in 2D Dirac semimetals
\cite{Throckmorton15}. After making RG analysis, Moon \emph{et al.}
showed that Coulomb interaction is relevant in 3D semimetals with
quadratic band-touching points and can induce unusual non-FL
behaviors \cite{Moon13}, which confirms early studies of Abrikosov
and Beneslavskii \cite{Abrikosov71, Abrikosov74}. In 3D anisotropic
Weyl semimetals, Coulomb interaction is irrelevant, as revealed in a
recent RG analysis of Yang \emph{et al.} \cite{Yang2014Nat.P} and a
pioneering work of Abrikosov \cite{Abrikosov72}. Theoretic studies
\cite{Lai15, Jian15} claimed that Coulomb interaction is marginally
irrelevant in double Weyl semimetals, and gives rise to logarithmic
like corrections to various observable quantities. More recently,
Huh \emph{et al.} \cite{Huh16} demonstrated that Coulomb interaction
is also irrelevant in nodal-line semimetals.

In addition to these semimetals, a new type of 2D semimetal, often
called semi-Dirac semimetal (SDSM), has also attracted considerable
theoretical and experimental interest \cite{Isobe2016PRL,
Moon2016SR, Lim2012PRL, Dora2013PRB, Delplace2013PRB, Huang2015PRB,
Pyatkovskiy2016PRB, Montambaux2009EPJB, Montambaux2009PRB,
Dietl2008PRL, Katayama2006JPSJ, Kobayashi2007JPSJ, Goerbig2008PRB,
Pardo2009PRL, Pardo2010PRB, Banerjee2009PRL, Delplace2010PRB,
Banerjee2012PRB, Ezawa2014NJP, Q.Liu2015NaL, Dolui2015SR,
Xiang2015SR, Gong2016PRB, Kim2015Sci}. Close to the band touching
points, the fermion dispersion is linear along one momentum
direction and quadratic along the other, so a suitable name for the
fermion of this system is semi-Dirac fermion (SDF). Possible
materials that are predicted to realize such SDFs include deformed
graphene \cite{Montambaux2009EPJB, Montambaux2009PRB, Dietl2008PRL},
compressed organic conductor $\alpha$-(BEDT-TTF)$_2$I$_3$
\cite{Katayama2006JPSJ, Kobayashi2007JPSJ, Goerbig2008PRB},
specifically designed TiO$_2$/VO$_2$ nanostructure
\cite{Pardo2009PRL, Pardo2010PRB, Banerjee2009PRL, Delplace2010PRB,
Banerjee2012PRB}, and black phosphorous subject to strain
\cite{Ezawa2014NJP, Q.Liu2015NaL, Dolui2015SR,Xiang2015SR,
Gong2016PRB, Kim2015Sci}. In the non-interacting limit, the fermion
DOS depends on energy $\omega$ in the form $\rho(\omega) \propto
\sqrt{\omega}$ for small values of $\omega$. The Coulomb interaction
is only partially screened, and has recently been showed to induce
non- or marginal FL behaviors over a wide range of energy scales
\cite{Isobe2016PRL}.

Disorder effects is another important research field in condensed
matter physics \cite{Lee85, Altshuler, Belitz94, Abrahams01,
Altland02, Sarma11, Kotov12}. In 3D ordinary metals, the electrons
become localized when the (non-magnetic) disorder is sufficiently
strong, leading to metal-insulator transition \cite{Lee85,
Belitz94}. The scaling theory of localization leads to a striking
conclusion that an arbitrarily weak disorder is able to cause weak
localization of electrons in a 2D metal \cite{Abrahams79, Lee85},
provided that Coulomb interaction is ignored. Adding a weak Coulomb
interaction results in analogous localization behavior in a
disordered Fermi liquid \cite{Altshuler}. It thus turns out that 2D
metallic state cannot exist in zero magnetic field. Surprisingly,
experiments have found compelling evidences for the existence of
metallic behaviors in certain 2D dilute dirty electron systems
\cite{Abrahams01}, which challenges the traditional view of 2D
localization. Although the microscopic mechanism of 2D
metal-insulator transition is still in debate, it is universally
believed that strong Coulomb interaction should play a critical role
\cite{Abrahams01}.

Similar to the Coulomb interaction, disorder also has different
effects in various semimetals comparing to normal metals. In Dirac
semimetals, backscattering is suppressed due to the chirality of
massless Dirac fermions \cite{Sarma11, Altland02, Castroneto09,
Kotov12}. Strong disorder may drive the system to undergo a quantum
phase transition from a ballistic state to a diffusive state by
generating a finite DOS at the Fermi level and a finite scattering
rate. Moreover, disorder appears in several different forms such as
random energy and random fermion mass, distinguished by their
coupling to fermions \cite{Ludwig1994, Altland02, Castroneto09,
Sarma11}. There are three types of most frequently studied disorders
\cite{Ludwig1994, Stauber2005PRB, Herbut08, Vafek08, Moon2014arXiv,
WangLiu14}, including random chemical potential, random mass, and
random gauge potential. The massless Dirac/Weyl fermions usually
display distinct behaviors when they couple to different types of
disorder \cite{WangLiu14, Ludwig1994, Stauber2005PRB, Herbut08,
Vafek08, Moon2014arXiv, Carpentier13}.

When Coulomb interaction and disorder are both present, they have
mutual influences on each other. Therefore, in many cases it is
necessary to study their interplay \cite{Lee85, Belitz94}.
Finkelstein \cite{Finkelstein84} has developed a RG scheme to study
this problem and found evidence for 2D metallic behaviors.
Unfortunately, the interaction parameter flows to infinity, which
invalidates the RG analysis based on the weak coupling expansion
\cite{Finkelstein84, Belitz94, Abrahams01}. However, if the system
contains $N$ species of fermions, it would be possible to carry out
RG calculations by making series expansion in powers of $1/N$
\cite{Finkelstein05}, which allows one to access the strong
interaction regime. In recent years, the interplay of Coulomb
interaction and disorder has been studied extensively in the
contexts of various semimetals \cite{WangLiu14, Stauber2005PRB,
Herbut08, Vafek08, Goswami2011PRL, Moon2014arXiv}, and found to
result in quantum phase transitions, non-FL behaviors, and other
unusual properties.

\begin{table}[pth]
\caption{The main RG results are briefly summarized in the Table.
Col. is the abbreviation for the term Coulomb interaction, Dis. for
disorder, RL for relevant, IR for irrelevant, MG for marginal, MR
for marginally relevant, MIR for marginally irrelevant, and P.T. for
phase transition. The parameter $\alpha_N = N\alpha$ characterizes
the effective strength of Coulomb interaction, $C_{g}$ is the
effective strength parameter for disorder, defined by
Eq.~(\ref{Eq.C_g}), with $C_{g0}$ being its bare value, and $a_c$ is
a critical strength given by Eq.~(\ref{Eq_a_c}).}\label{Tab_summary}
\begin{center}
\begin{tabular}{|*{4}{c|}}
\hline
\multirow{2}*{\backslashbox{Dis.}{Col.}}
&\,\,None\,\,&Strong&Weak\\
&\,\,($\alpha_N=0$)\,\,&($\alpha_N\gg1$)&($0<\alpha_N\ll1$)\\
\hline \cline{1-3}
&&\,$C_g$: IR $\xrightarrow[P.T.]{C_{g0} > a_c}$ RL \,&\\
Chemical&MR&&\\
&&$\alpha$: IR $\xrightarrow[P.T.]{C_{g0} > a_c}$ RL&\\
\cline{1-3}
&&&MIR \\
Mass&MIR&$C_g$: MR&\\
&&$\alpha$ : RL&\\
\cline{1-3}
&&&\\
Gauge&MG&$C_g$: RL&\\
&&$\alpha$ : RL&\\
\hline
\end{tabular}
\end{center}
\end{table}

In this paper, we study the interplay of long-range Coulomb
interaction and non-magnetic disorder in an 2D SDF system, focusing
on their mutual influences and the physical consequences produced by
such an interplay. Following Isobe \emph{et al.}
\cite{Isobe2016PRL}, we will treat the Coulomb interaction between
SDFs by employing a $1/N$ expansion and perform RG calculations in
both the weak and strong coupling limits, corresponding respectively
to small and large values of Coulomb interaction parameter $\alpha$
defined below in Eq.~(\ref{Eq.alpha}). However, disorders are
assumed to be weak. We consider the aforementioned three types of
disorders and show that they may significantly change the role
played by the Coulomb interaction. On the other hand, the role of
disorder can also be substantially altered by the Coulomb
interaction. The main RG results are briefly summarized in
Table~\ref{Tab_summaey}, and explained in more detail in the
following.

(1) In the case of random chemical potential, the effective disorder
parameter $C_g$ has an unstable fixed point $C_g^{\star}$. If the
Coulomb interaction parameter $\alpha_N \gg 1$ and the initial value
of $C_g$ is larger than $C_g^{\star}$, the Coulomb interaction and
random chemical potential promote each other, with $\alpha_N$ and
$C_g$ growing indefinitely as the energy is lowered. This behavior
most likely signals the happening of a quantum phase transition
between semimetal and certain insulator. In the weak coupling limit
with $\alpha_N \ll 1$, Coulomb interaction and random chemical
potential cancel each other, with $\alpha_N$ and $C_g$ both flowing
to the Gaussian fixed point. In different energy regimes, the system
exhibits non- or marginal FL behaviors.

(2) In the case of random mass, the low-energy behaviors of the
system depend sensitively on the initial values of $\alpha_N$ and
$C_g$. In the strong coupling limit $\alpha_N \gg 1$, there is a
stable infrared fixed point for $C_g$, at which the system displays
non-FL behaviors. Random mass, being marginally irrelevant in the
non-interacting limit, becomes marginally relevant due to the strong
Coulomb interaction. In the weak coupling limit, there is a stable
Gaussian fixed point and an unstable finite fixed point
$\left(\alpha^{\star}, C_g^{\star}\right)$. Beyond
$\left(\alpha^{\star}, C_g^{\star}\right)$, the parameters
$\alpha_N$ and $C_g$ increase at low energies. Below
$\left(\alpha^{\star}, C_g^{\star}\right)$, both $\alpha_N$ and
$C_g$ flow to zero at low energies and the low-energy behavior is
essentially governed by the Coulomb interaction.

(3) Random gauge potential and Coulomb interaction promote each
other, and both $\alpha_N$ and $C_g$ flow to strong coupling regime
if $\alpha_N \gg 1$. Similar to the case of random chemical
potential, the SDSM might enter into certain insulating phase. In
the weak coupling limit $\alpha_N \ll 1$, there are also two fixed
points for $\alpha_N$ and $C_g$: a stable Gaussian fixed point, and
an unstable finite fixed point. In the latter case, the results are
analogous to the other two types of disorder, and Coulomb
interaction governs the system in the low-energy regime, with
disorder playing an unimportant role.

Unfortunately, it is technically difficult to access the
intermediate coupling regime of Coulomb interaction at present. It
thus remains unclear how to reconcile the results separately
obtained in the weak and strong coupling limits. Moreover, though it
seems reasonable to speculate that the ground state of the strong
coupling regime is insulating, its nature is unknown and further
study is required.

The rest of the paper is organized as follows. We present the whole
model in Sec.~\ref{Sec_model} and perform detailed RG calculations
in the presence of three types of disorders and Coulomb interaction
in Sec.~\ref{Sec_full_RG}. We ignore Coulomb interaction and analyze
the impact of disorders on the low-energy behaviors of 2D SDFs in
Sec.~\ref{Sec_clean_analysis}. In Sec.~\ref{Sec_Interplay_soultion},
we solve the RG equations after taking into account the mutual
influences between Coulomb interaction and disorder, and use the
solutions to examine the roles played by Coulomb interaction and
disorder are affected by each other. We summarize the results and
briefly discuss the physical implications of our results in
Sec.~\ref{Sec_summary}.

\section{Effective model}\label{Sec_model}

We consider 2D SDF in the presence of both Coulomb interaction and
disorder. The effective action is written in the momentum-energy
space as follows
\begin{eqnarray}
S = S_f + S_b + S_g + S_{\mathrm{dis}},\label{Eq.S1}
\end{eqnarray}
where
\begin{eqnarray}
S_f &=& \!\sum_{i=1}^{N}\int \frac{d\omega}{2\pi}
\frac{d^2k}{(2\pi)^2} \psi_i^{\dag}(-i\omega + Ak_{x}^2\tau_x +
vk_y\tau_y) \psi_i,\label{Eq.S_F}\\
S_b &=& \frac{1}{2}\int \frac{d\omega}{2\pi} \frac{d^3k}{(2\pi)^2}
\phi^{\dag}(\omega, \mathbf{k})(k_{x}^2 + k_{y}^2+k_{z}^2)
\phi(\omega, \mathbf{k}),\label{Eq.S_B}\\
S_g &=& \sum_{i=1}^{N}\int \frac{d\omega}{2\pi}
\frac{d\omega'}{2\pi}\frac{d^2k}{(2\pi)^2}
\frac{d^2k'}{(2\pi)^2}\psi_i^{\dag}(\omega, \mathbf{k})\nonumber\\
&&\times\left[ig\phi(\omega-\omega',\mathbf{k-k'})\right]
\psi_i(\omega',\mathbf{k'}).\label{Eq.S_g}
\end{eqnarray}
Here, $\psi_i$ describes a two-component spinor field with the
subscript $i = 1, ..., N$ labeling the species of fermions. The
fermions have an anisotropic dispersion, as shown by the free action
term $S_f$, where $A$ is the inverse of mass along the $x$-axis and
$v$ an effective velocity along the $y$-axis \cite{Isobe2016PRL}.
Both $\tau_x$ and $\tau_y$ are the standard Pauli matrices. The
instantaneous Coulomb interaction is described by the Yukawa
coupling term $S_g$ between spinor fields $\psi_i$ and a bosonic
field $\phi$ that is introduced by performing a Hubbard-Stratonovich
transformation \cite{Isobe2016PRL}. The parameter $g$ characterizes
the strength of Yukawa coupling, and is given by $g =
e/\sqrt{\varepsilon}$, where $e$ is the electric charge and
$\varepsilon$ dielectric constant. It is more convenient to define a
dimensionless coupling parameter $\alpha$, which appears as a whole
throughout all the perturbative calculations, to denote the ratio
between Coulomb potential energy $E_{\mathrm{c}}\sim A^{-1}v g^2$
and fermion kinetic energy $E_{\mathrm{k}}\sim A^{-1}v^2$. One can
easily find that
\begin{eqnarray}
\alpha \equiv \frac{E_{\mathrm{c}}}{E_{\mathrm{k}}} =
\frac{g^2}{v}.\label{Eq.alpha}
\end{eqnarray}
As explained in Ref.~\cite{Isobe2016PRL}, the bosonic field $\phi$
is defined in 3D space, whereas the Dirac fermions are confined to
the $(x,y)$-plane. Integration over $k_z$ gives rise to the Coulomb
potential: $D_0(\mathbf{q})\propto1/|\mathbf{q}|$.

From the free action terms $S_f$ and $S_b$, it is straightforward to
obtain the free propagators of $\psi$ and $\phi$:
\begin{eqnarray}
G_0(\omega, \mathbf{k}) &=& \frac{1}{-i\omega + Ak_{x}^2\tau_x +
vk_y\tau_y},\\
D_0(\mathbf{q}) &=& \int\frac{dq_{z}}{2\pi}\frac{1}{q_{x}^2 + q_y^2
+ q_{z}^2} \nonumber \\
&=& \frac{1}{2\sqrt{q_{x}^2 + q_y^2}}.
\end{eqnarray}
Due to the anisotropy in the fermion dispersion, we need to perform
unusual scaling transformations. In the non-interacting limit, the
energy and momenta are generically rescaled as
\begin{eqnarray}
\tilde{\omega} = b^z\omega\,\,\tilde{k}_{x} = bk_x,\,\,\tilde{k}_{y}
= b^{z_1}k_y.\label{Eq.Scaling}
\end{eqnarray}
The scaling dimensions are defined by $[\omega] = z$, $[k_x] = 1$,
and $[k_y] = z_1$. Based on the free action term $S_f$, it is easy
to verify that $[A] = z-2$ and $[v] = z-z_1$. After including the
corrections induced by Coulomb interaction, the parameter $A$ and
$v$ will be renormalized and exhibit unusual dependence on varying
energy scale.

The effects of Coulomb interaction on the 2D SDFs have been
systematically analyzed by Isobe \emph{et al.} in a recent work
\cite{Isobe2016PRL}. We will go one step further and include
quenched disorder into the interacting SDF system. To make an
unbiased analysis, we shall treat Coulomb interaction and
fermion-disorder coupling on an equal footing, and study their
mutual influences by means of RG approach.

\begin{figure}[htbp]
\center
\includegraphics[width=3.35in]{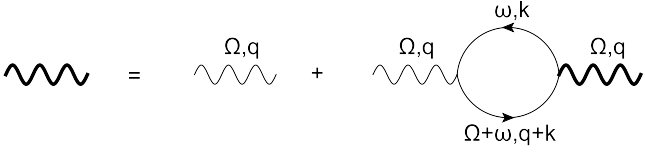}
\vspace{-0.20cm} \caption{One-loop Feynman diagram for dynamically
screened Coulomb interaction. The solid line represents the free
propagator of fermions, thin wavy line represents the bare Coulomb
interaction, and thick wavy line represents the screened Coulomb
interaction. The dynamical screening effects are embodied by the
fermion loop to the leading order of $1/N$ expansion.}
\label{Fig_polarization}
\end{figure}
\begin{figure}[htbp]
\center \subfigure{
\includegraphics[width=2.8in]{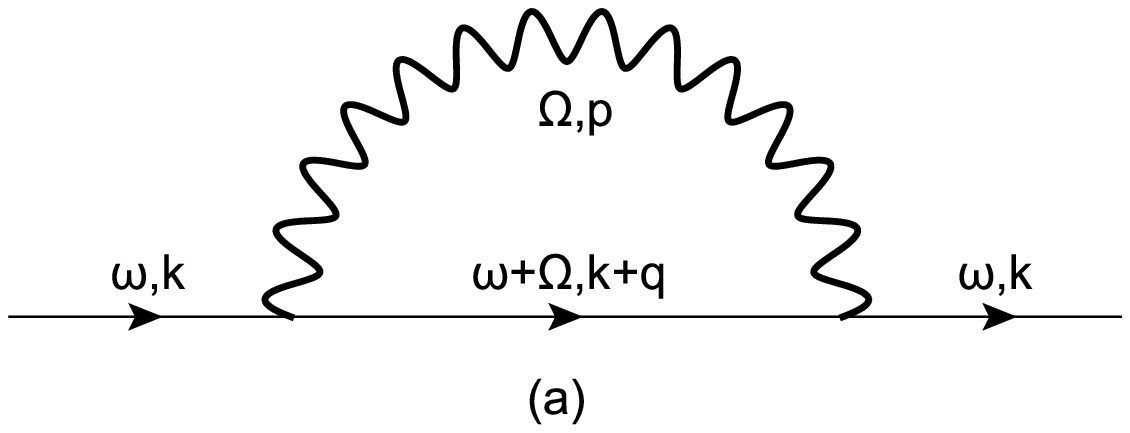}\label{Fig_Coulomb_self_en}}\vspace{0.5cm}
\subfigure{
\includegraphics[width=2.8in]{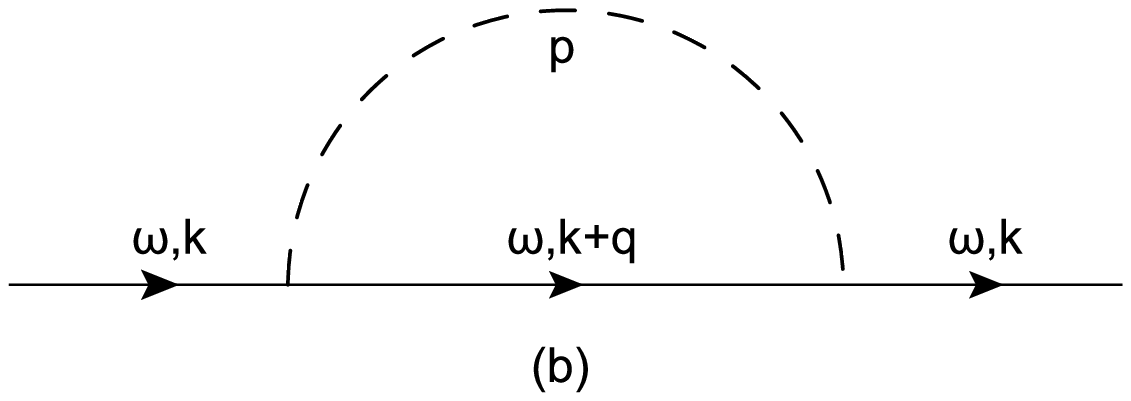}\label{Fig_disorder_self_en}}
\caption{Leading order corrections to the fermion self-energy due to
(a) Coulomb interaction; (b) disorder. In diagram (b), the dashed
line stands for disorder, corresponding to the mean value $\langle
V(\mathbf{x})V(\mathbf{x}')\rangle$, and the coupling between
fermion and disorder is represented by the intersection of solid and
dashed lines.} \label{Fig_Fermion_SE}
\end{figure}

The action for the fermion-disorder coupling that needs to be added
to the system has the following form \cite{Ludwig1994, Nersesyan95,
Altland02, Stauber2005PRB, Herbut08, Vafek08, Wang2011PRB,
WangLiu14},
\begin{eqnarray}
S_{\mathrm{dis}} = \sum_{i=1}^{N}v_{\Gamma}\int d^2 \mathbf{x} dt\,
\psi_{i}^\dagger(\mathbf{x})\Gamma \psi_{i}(\mathbf{x})
V(\mathbf{x}),
\end{eqnarray}
where $v_{\Gamma}$ is a coupling constant, and the function
$V(\mathbf{x})$ is a quenched, random variable behaving as a
Gaussian white noise, defined by
\begin{eqnarray}
\langle V(\mathbf{x})\rangle = 0,\qquad \langle
V(\mathbf{x})V(\mathbf{x}')\rangle = \Delta
\delta^2(\mathbf{x}-\mathbf{x}').\label{Eq_def_dis}
\end{eqnarray}
Here, a dimensionless variance $\Delta$ is introduced to
characterize the strength of random potential.

We will consider three different types of disorder classified by the
different forms of the matrix $\Gamma$. In particular, $\Gamma =
\textmd{I}$ for random chemical potential, $\Gamma = \tau_{z}$ for
random mass, and $\Gamma = (\tau_{x},\tau_{y})$ for random gauge
potential. The RG calculations are most conveniently carried out in
the energy-momentum space. After making a Fourier transformation, we
rewrite $S_{\mathrm{dis}}$ as an integration over energy and
momentum in the following form:
\begin{eqnarray}
S_{\mathrm{dis}} &=& \sum_{i=1}^{N}\int\frac{d\omega}{2\pi}
\frac{d^{2}\mathbf{k}}{(2\pi)^2}\frac{d^{2}\mathbf{k'}}{(2\pi)^2}
\psi_{i}^\dagger(i\omega,\mathbf{k})(v_{\Gamma}\Gamma)
\psi_{i}(i\omega,\mathbf{k'})\nonumber \\
&& \times V(\mathbf{k-k'}).
\end{eqnarray}
This action will be analyzed together with
Eqs.~(\ref{Eq.S_F}) - (\ref{Eq.S_g}). In the following calculations,
we choose to treat disorder using the ordinary perturbative
expansion method \cite{Lee85, Stauber2005PRB}. Within this
formalism, the coupling between fermions and random potential is
represented by the diagrams presented in
Fig.~\ref{Fig_disorder_self_en},
Fig.~\ref{Fig_Coulomb_vertex_disorder}, and
Fig.~\ref{Fig_all_disorder_vertex}.

\section{Full RG eqations to the leading order of perturbative expansion}
\label{Sec_full_RG}

In this section, we derive the full set of RG equations for all the
parameters appearing in the whole action. The calculations are
carried out to the leading order of perturbative expansion. To make
our analysis self-contained, we first briefly summarize the main
results already obtained in Ref.~\cite{Isobe2016PRL} in
Sec.~\ref{Sec_clean_correction}, listing all the formulae to be used
in the following calculations. The corrections induced by disorders
are calculated in Sec.~\ref{Sec_one_loop_disorder}. Based on these
results, we deduce all the RG equations in Sec.~\ref{Sec_full_RGs}.
The solutions of RG equations along with their physical implications
will be presented in the next section.

\subsection{Clean limit}\label{Sec_clean_correction}

To the leading order of $1/N$ expansion, the diagram for the
polarization function is shown in Fig.~\ref{Fig_polarization}, which
is expressed as
\begin{eqnarray}
\Pi_1(\Omega,\mathbf{q}) &=& -N\int\frac{d\omega}{2\pi}
\frac{d^2k}{(2\pi)^2} \mathrm{Tr}[G_{0}(\omega,\mathbf{k})
\nonumber \\
&&\times G_{0}(\omega + \Omega,\mathbf{k} + \mathbf{q})].
\end{eqnarray}
It is hard to obtain an analytical expression for this polarization.
As shown by Isobe \emph{et al.} \cite{Isobe2016PRL}, it can be well
approximated by the following \emph{ansatz}
\begin{eqnarray}
\Pi_1(\Omega, \mathbf{q})&=&-\alpha_N\Big[\frac{d_{x}A^{1/2}
q_{x}^2}{(\Omega^2+cA^2q_{x}^4+v^2q_{y}^2)^{1/4}}
\nonumber\\&+&\frac{d_{y}A^{-1/2}v^2q_{y}^2}{(\Omega^2 + cA^2
q_{x}^4 + v^2 q_{y}^2)^{3/4}}\Big],
\end{eqnarray}
where
\begin{eqnarray}
\alpha_N &\equiv& N \alpha,\,\,\,\, c = \left(\frac{2}{\sqrt{\pi}}
\frac{\Gamma(3/4)}{\Gamma(9/4)}\right)^4,\nonumber \\
d_{x} &=& \frac{1}{8\sqrt{\pi}}
\frac{\Gamma(3/4)}{\Gamma(9/4)},\,\,\,\,d_{y} =
\frac{1}{8\sqrt{\pi}}\frac{\Gamma(5/4)}{\Gamma(7/4)}.
\end{eqnarray}
After including the dynamical screening effects, as diagrammatically
depicted in Fig.~\ref{Fig_polarization}, we use Dyson equation to
obtain the inverse of effective propagator of $\phi$:
\begin{eqnarray}
D^{-1}(\Omega, \mathbf{q}) &=& D_{0}^{-1}(\Omega,
\mathbf{q}) - \Pi_1(\Omega,\mathbf{q}) \nonumber \\
&=& 2\sqrt{q_{x}^2 + q_y^2} + \alpha_{N}\Big[\frac{d_{x}
A^{1/2}q_{x}^2}{(\Omega^2 +cA^2 q_{x}^4 + v^2 q_{y}^2)^{1/4}}
\nonumber \\
&&+\frac{d_{y}A^{-1/2}v^2 q_{y}^2}{(\Omega^2 + cA^2 q_{x}^4 + v^2
q_{y}^2)^{3/4}}\Big].\label{Eq.boson_prog}
\end{eqnarray}

\begin{figure}[htbp]
\center \subfigure{
\includegraphics[width=2.45in]{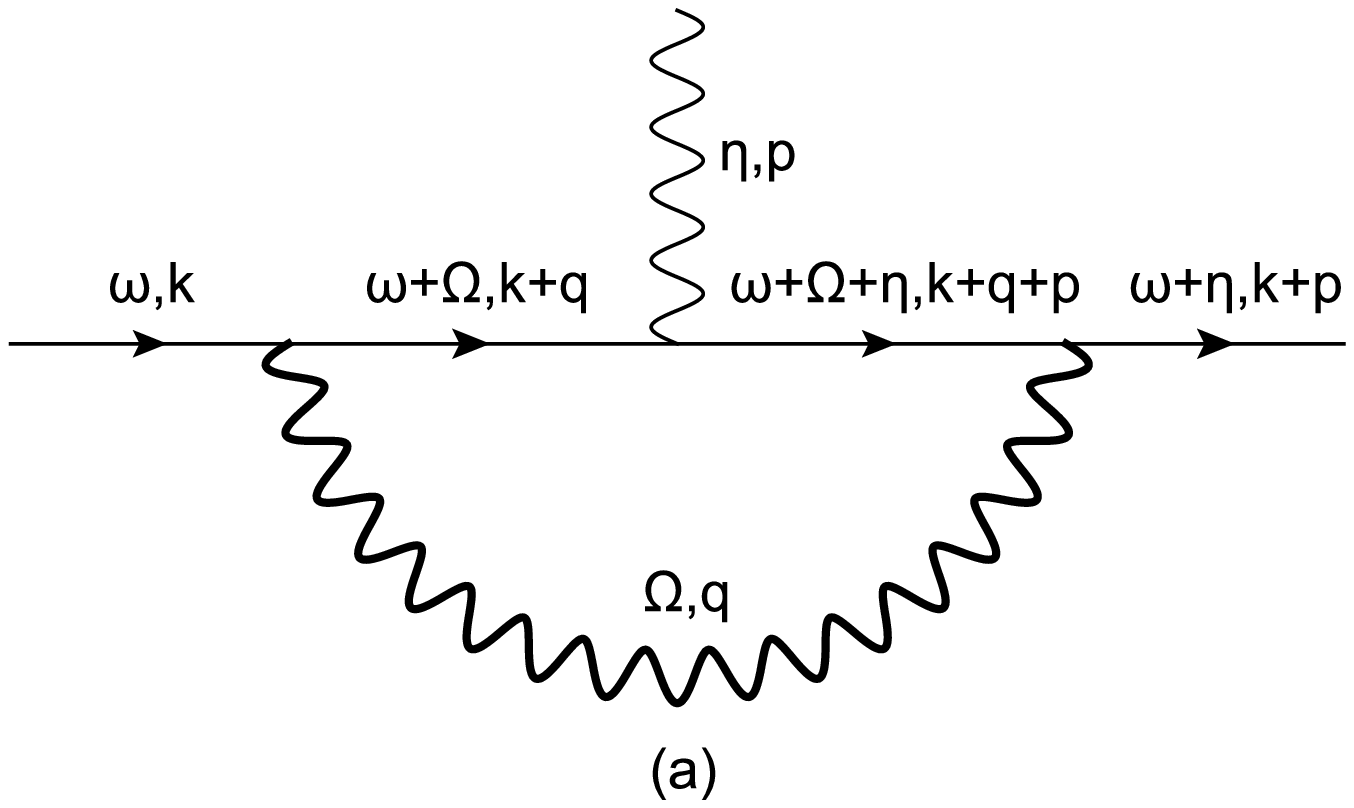}
\label{Fig_Coulomb_vertex_Coulomb}}\\
\subfigure{
\includegraphics[width=2.45in]{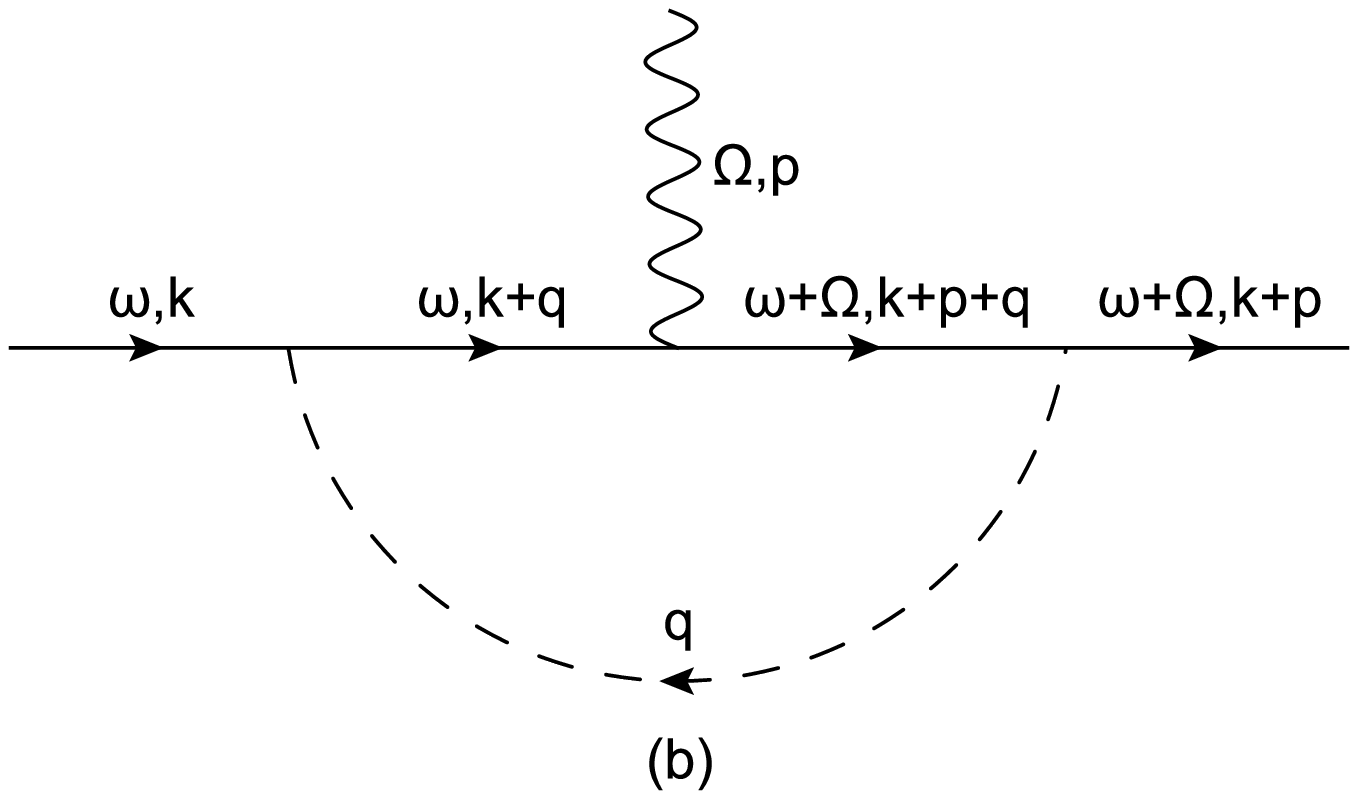}
\label{Fig_Coulomb_vertex_disorder}} \caption{One-loop Coulomb
vertex correction due to (a) Coulomb interaction, (b)
disorder.}\label{Fig_Coul_vert}
\end{figure}

According to Fig.~\ref{Fig_Coulomb_self_en}, the fermion self-energy
correction due to the Coulomb interaction can be computed as follows
\begin{eqnarray}
\Sigma_{1}(i\omega,\mathbf{k}) &=& -g^2\int\frac{d\Omega
d^2q}{(2\pi)^3}G_{0}(\omega+\Omega,\mathbf{k+q})D(\Omega,
\mathbf{q}) \nonumber \\
&=& \Sigma_{\omega}i\omega - \Sigma_{k_{x}}Ak_{x}^2\tau_{x} -
\Sigma_{k_{y}}vk_{y}\tau_{y},\label{Eq.self_en}
\end{eqnarray}
where
\begin{eqnarray}
\Sigma_{\omega}=\gamma_z l,\,\,\,\,\Sigma_{k_{x}} =
\Sigma_{\omega}+\gamma_A l,\,\,\,\,\Sigma_{k_{y}} =
\Sigma_{\omega}+\gamma_v l.\label{Eq.self_en_ana}
\end{eqnarray}
Here, a varying length scale $l = \log(\Lambda/\mu)$ is defined and
will be used later.

To compute $\Sigma_{1}(i\omega,\mathbf{k})$, we need to substitute
Eq.~(\ref{Eq.boson_prog}) into Eq.~(\ref{Eq.self_en}). However,
$D(\Omega,\mathbf{q})$ is formally very complicated, so it is hard
to get an analytic expression for $\Sigma_{1}(i\omega,\mathbf{k})$.
As demonstrated in Ref.~\cite{Isobe2016PRL}, it is convenient to
consider two limits $\alpha_N \gg 1$ and $\alpha_N \ll 1$
separately. In a given Dirac fermions system, the fermion flavor is
fixed at certain value, but the parameter $\alpha$ can be tuned by
changing the surrounding environment or other variables. Therefore,
these two limits can be achieved by tuning $\alpha$ to take a large
or small value, corresponding to strong or weak coupling limit
respectively. In the strong coupling limit with $\alpha_N \gg 1$,
$\Pi_1(\Omega,\mathbf{q})$ is much more important than
$D_0^{-1}(\Omega,\mathbf{q})$. Conversely,
$D_0^{-1}(\Omega,\mathbf{q})$ is much more important than
$\Pi_1(\Omega,\mathbf{q})$ in the weak coupling limit with $\alpha_N
\ll 1$. In both cases, $D^{-1}(\Omega,\mathbf{q})$ is simple enough
to lead to an analytical expression for
$\Sigma_{1}(i\omega,\mathbf{k})$. It was found in
Ref.~\cite{Isobe2016PRL} that the anomalous exponents defined in
Eq.~(\ref{Eq.self_en}) are
\begin{eqnarray}
\gamma_z = \frac{\sqrt{15}
\log(\alpha_{N})}{\pi^{(3/2)}N},\,\,\,\gamma_A =
\frac{0.1261}{N},\,\,\,\gamma_v = \frac{0.3625}{N}
\end{eqnarray}
in the strong coupling limit, and
\begin{eqnarray}
\gamma_z = \frac{3\alpha_N}{8\pi^2N},\,\,\,\,\gamma_A =
\frac{\alpha_N|\ln\alpha_N|}{2\pi^2N},\,\,\,\,\gamma_v =
\frac{\alpha_N}{4\pi^2N}
\end{eqnarray}
in the weak coupling limit.

\begin{figure}[htbp]
\center \subfigure{
\includegraphics[width=2.58in]{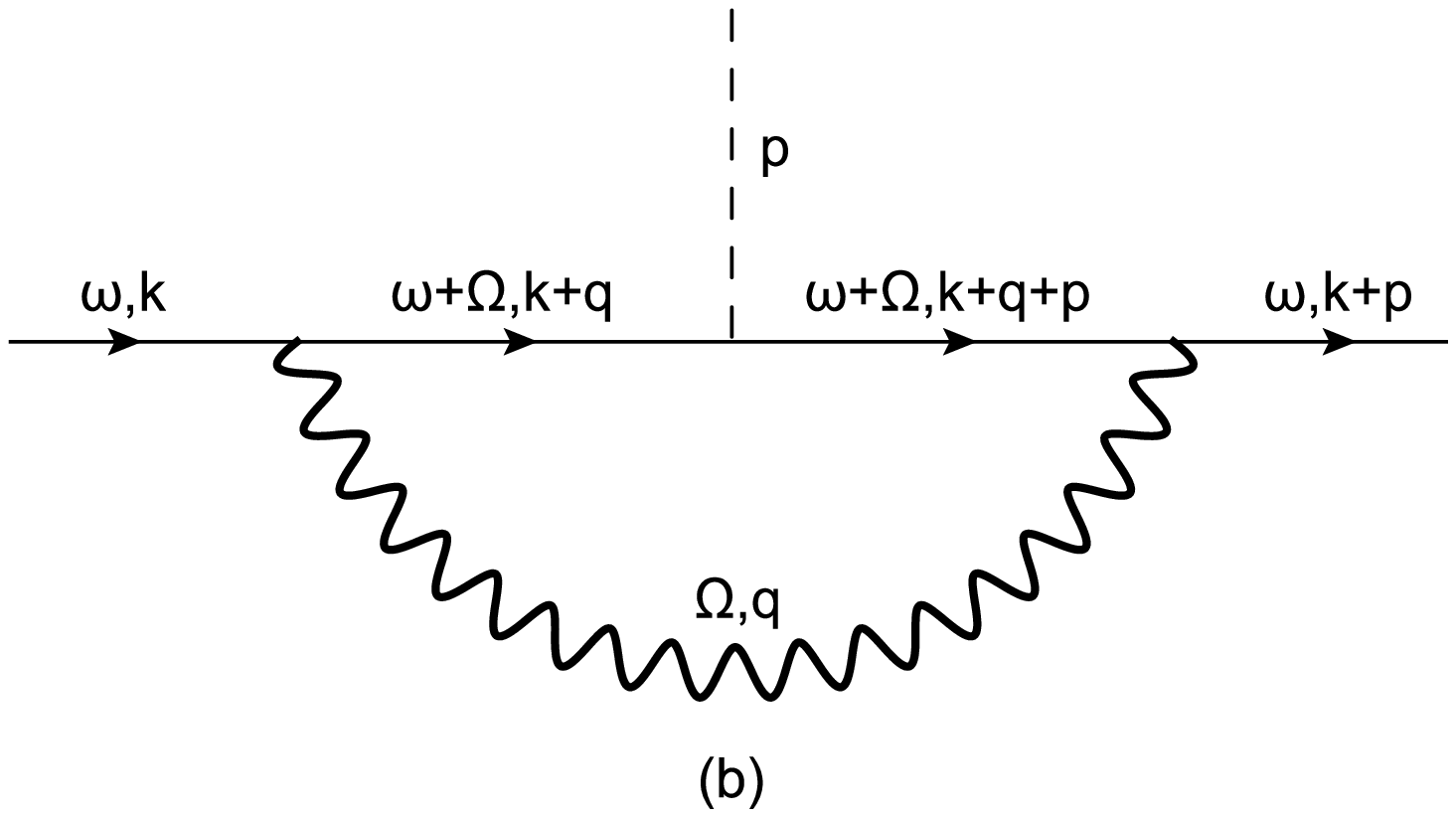}
\label{Fig_dis_vertex_Coulomb}}\\
\subfigure{
\includegraphics[width=2.58in]{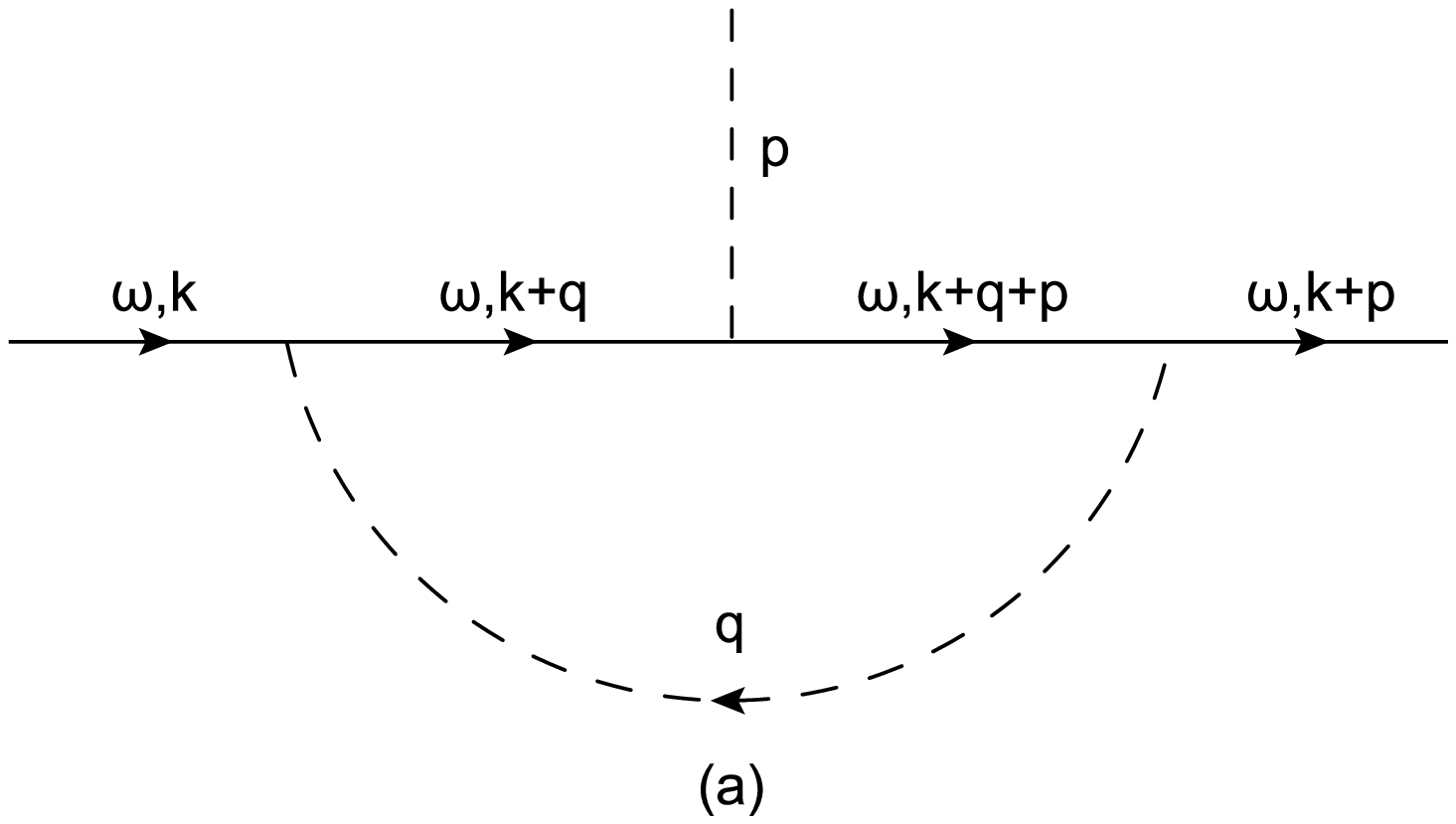}
\label{Fig_disorder_vertex}}\caption{One-loop fermion-disorder
vertex correction due to (a) Coulomb interaction, (b)
disorder.}\label{Fig_all_disorder_vertex}
\end{figure}

All the vertex corrections for Coulomb interaction are shown in
Fig.~\ref{Fig_Coul_vert}. As demonstrated in
Ref.~\cite{Isobe2016PRL}, at vanishing external momenta and energy,
the vertex correction due to Coulomb correction shown in
Fig.~\ref{Fig_Coulomb_vertex_Coulomb} is \cite{Isobe2016PRL}
\begin{eqnarray}
\delta g_{1} &=& -ig^3\int\frac{d\Omega d^2q}{(2\pi)^3}
G_{0}(\Omega,\mathbf{q})G_{0}(\Omega,\mathbf{q})D(\Omega,
\mathbf{q}) \nonumber \\
&\equiv& (ig)\Sigma_{\omega}.
\end{eqnarray}

\subsection{Corrections with Fermion-disorder interaction}\label{Sec_one_loop_disorder}

The leading correction to the fermion self-energy due to
fermion-disorder interaction is shown in
Fig.~\ref{Fig_disorder_self_en}, which is computed as follows
\begin{eqnarray}
\Sigma_{\mathrm{dis}}(i\omega) &=& \Delta v_{\Gamma}^{2}
\int\frac{d^2\mathbf{q}}{(2\pi)^2}\Gamma
G_{0}(i\omega,\mathbf{q})\Gamma \nonumber \\
&=& i\omega\frac{\Delta v_{\Gamma}^{2}}{2\pi v^2} \frac{1}{\bar{A}}l
+ \frac{\Delta v_{\Gamma}^{2}}{2\pi v} \Lambda\Gamma\tau_x \Gamma,
\label{Eq.SE_dis_b}
\end{eqnarray}
where $\bar{A} \equiv \frac{A\Lambda}{v}$. Similar to the case
studied in Ref.~\cite{Isobe2016PRL}, the self-energy contains a
constant contribution $\Sigma_{\mathrm{dis}}(0,0) = \frac{\Delta
v_{\Gamma}^{2}}{2\pi v}\Lambda\Gamma\tau_x \Gamma$, which is not
divergent in the infrared region and thus can be safely neglected.
Moreover, $\Sigma_{\mathrm{dis}}$ does not depend on momenta since
the quenched disorders are static. It is also independent of the
specific sort of the disorder. To capture these two important
features, we rewrite the self-energy as
\begin{eqnarray}
\Sigma_{\mathrm{dis}} = i\omega C_g l,\label{Eq.SE_dis_f}
\end{eqnarray}
where a dimensionless parameter is defined to characterize the
effective strength of disorder:
\begin{eqnarray}
C_g \equiv \frac{\Delta v_{\Gamma}^{2}}{2\pi
v^2}\frac{1}{\bar{A}}.\label{Eq.C_g}
\end{eqnarray}

We next consider the corrections to the fermion-disorder vertex,
which receive contributions from both Coulomb interaction and
fermion-disorder interaction, represented by the diagrams shown in
Fig.~\ref{Fig_all_disorder_vertex}. At zero external momenta, the
vertex correction due to disorder, corresponding to
Fig.~\ref{Fig_disorder_vertex}, is
\begin{eqnarray}
V_{\mathrm{dis}} = \Delta v_{\Gamma}^{2}\int
\frac{d^2\mathbf{q}}{(2\pi)^2}\Gamma G_0(i\omega,\mathbf{q})
v_\Gamma\Gamma G_0(i\omega,\mathbf{q})\Gamma,
\end{eqnarray}
which yields
\begin{eqnarray}
V_{\mathrm{dis}} &=&C_D(v_\Gamma \Gamma) l.\label{Eq.Cp_dis}
\end{eqnarray}
where $C_D = C_g$ for random chemical potential, $C_D= 0$ for both
of the two components of random gauge potential, and $C_D= -C_g$ for
random mass.

According to Fig.~\ref{Fig_dis_vertex_Coulomb}, at zero external
momenta and energy, the disorder vertex correction due to Coulomb
interaction is given by
\begin{eqnarray}
V_{\mathrm{c}} = -g^2\int\frac{d\Omega d^2\mathbf{q}}{(2\pi)^3}
G_{0}(\Omega,\mathbf{q})(v_{\Gamma}\Gamma)
G_{0}(\Omega,\mathbf{q})D(\Omega,\mathbf{q}),\label{Eq.dis_vert_C}
\end{eqnarray}
Here, we borrow some definitions from Ref.~\cite{Isobe2016PRL} and
then obtain
\begin{eqnarray}
V_{\mathrm{c}} = (v_{\Gamma}\Gamma)
\left(\Sigma_{\omega}+\gamma_{D}l\right).\label{Eq.CP_Coul}
\end{eqnarray}
In the strong coupling limit, $\gamma_{D} = 0$ for random chemical
potential, $\gamma_{D} = 1.1564/N$ for random mass, $\gamma_{D} =
0.7939/N$ and $\gamma_{D}=0.3625/N$ for the $\Gamma=\tau_x$- and
$\Gamma=\tau_y$-component of random gauge potential respectively. In
the weak coupling limit, $\gamma_{D} = 0$ for random chemical
potential, $\gamma_{D} = 13\alpha_N/4\pi^2N$ for random mass,
$\gamma_{D} = 3\alpha_N/\pi^2N$ for the $\tau_x$-component and
$\gamma_{D}=\alpha_N/4\pi^2N$ for the $\tau_y$-component of random gauge
potential.


Moreover, disorder will produce another correction to Coulomb
interaction vertex, as shown in
Fig.~\ref{Fig_Coulomb_vertex_disorder}. This correction will be
calculated here to illustrate that $g$ is actually not renormalized
within the framework of Wilsonian RG scheme. According to
Fig.~\ref{Fig_Coulomb_vertex_disorder}, it is easy to get
\begin{eqnarray}
\delta g_2 &=& \Delta
v_{\Gamma}^{2}\int\frac{d^2\mathbf{q}}{(2\pi)^2} \Gamma
G_0(\mathbf{q})(ig)G_0(\mathbf{q})\Gamma \nonumber \\
&=& (ig)C_gl.\label{Eq.Coul_vert_dis}
\end{eqnarray}

\subsection{Derivation of RG equations}\label{Sec_full_RGs}

To perform RG analysis, it is convenient to define
\begin{eqnarray}
&&\tilde{\psi} = \sqrt{Z_{\psi}}\psi, \,\,\,\, \tilde{\phi} =
\sqrt{Z_{\phi}}\phi,\,\,\,\,\tilde{A}=Z_{A}A, \nonumber \\
&&\tilde{v} = Z_{v}v,\,\,\,\,\tilde{g} = Z_{g}g,
\,\,\,\,\tilde{v}_{\Gamma} = Z_{\Gamma}v_{\Gamma}.
\end{eqnarray}
Before applying rescaling transformation to the effective action, we
need to first determine how the random potential transforms under
the scaling given by Eq.~(\ref{Eq.Scaling}). In the spirit of RG
theory \cite{Shankar1994RMP}, to specify how a field operator
transforms, the standard method is to require that its kinetic term
remains invariant. However, the random potential $V(\mathbf{x})$
does not have its own kinetic term. To proceed, we write the
Gaussian white-noise distribution in the momentum space as
\begin{eqnarray}
\langle V(\mathbf{k}_{1}) V(\mathbf{k}_{2})\rangle &=&
\Delta\int\frac{d^2\mathbf{x}}{(2\pi)^{2}}
e^{i(\mathbf{k}_{1}+\mathbf{k}_{2})\cdot\mathbf{x}}\nonumber \\
&=&\Delta b^{z_1 + 1}\int\frac{d^2\tilde{\mathbf{x}}}{(2\pi)^{2}}
e^{i(\tilde{\mathbf{k}}_{1} +
\tilde{\mathbf{k}}_{2})\cdot\tilde{\mathbf{x}}}\label{Eq.res}.
\end{eqnarray}
By requiring the disorder distribution Eq.~(\ref{Eq.res}) to be
invariant under scaling transformations, we find that
\begin{eqnarray}
\tilde{V}(\tilde{\mathbf{k}})=V(\mathbf{k})b^{\frac{-1-z_1}{2}},
\end{eqnarray}
which will be used to carry out RG scaling transformations.

Based on the above results, we eventually obtain the following
complete set of flow equations:
\begin{eqnarray}
\frac{d\ln A}{dl} &=& \gamma_A - C_g + z - 2,\label{Eq.RG_A}\\
\frac{d\ln v}{dl} &=& \gamma_v - C_g + z - z_{1},\label{Eq.RG_v}\\
\frac{d\ln g}{dl} &=& \frac{z -
z_1}{2},\label{Eq.RG_g} \\
\frac{d\ln v_{\Gamma}}{dl} &=& \gamma_{D} + C_D - C_g + z -
\frac{1+z_1}{2}.\label{Eq.RG_vGamma}
\end{eqnarray}
As explained in Ref.~\cite{Isobe2016PRL}, the full vertex of Coulomb
interaction is given by the product $g Z_f$, which is guaranteed by
charge conservation to be unrenormalized. This property is reflected
in Eq.~(\ref{Eq.RG_g}), in which we use $g$ to represent the product
$g Z_f$ for notational simplicity.

In the above RG equations, the flowing parameter all have a finite
scaling dimension. To make a scaling-independent analysis of the
effects of Coulomb interaction and disorder, we need to derive flow
equations of dimensionless parameters. Using Eq.~(\ref{Eq.RG_v}) and
Eq.~(\ref{Eq.RG_g}), we obtain the flow equation for the Coulomb
interaction parameter
\begin{eqnarray}
\frac{d \ln \alpha_N}{dl} = C_g - \gamma_v,\label{Eq.RG_alpha}
\end{eqnarray}
where $\alpha_N$ is defined by the unrenormalized vertex $g$. For
fermion-disorder coupling parameter $C_g$, the corresponding flow
equation is
\begin{eqnarray}
\frac{d\ln C_g}{dl} = 2\gamma _{D} - \gamma_A - \gamma_v + 2C_D.
\label{Eq.RG_C_g}
\end{eqnarray}
We now have already obtained the RG equations for all the model
parameters. It is important to stress that the parameters
$\gamma_z$, $\gamma_A$, $\gamma_v$, and $\gamma_D$ take different
values in the strong and weak coupling limits, which will be
explained whenever necessary below.

By setting $C_g=0$ and $ z = z_1 = 2$, Eq.~(\ref{Eq.RG_A}),
Eq.~(\ref{Eq.RG_v}), and Eq.~(\ref{Eq.RG_alpha}) recover the RG
equations for $A$, $v$, and $\alpha$ previously obtained in
Ref.~\cite{Isobe2016PRL} in the clean limit. Isobe \emph{et al.}
\cite{Isobe2016PRL} have studied the effects of Coulomb interaction
on various observable quantities by performing a RG analysis, and
revealed that over a wide range of energies the Coulomb interaction
is a relevant perturbation in the strong coupling limit. In
particular, Coulomb interaction generates two anomalous exponents
$\gamma_A$ and $\gamma_v$ for parameters $A$ and $v$, respectively.
One can verify that $A(l)$ and $v(l)$ increase exponentially with
growing $l$, whereas the quasiparticle residue $Z_f(l)$ decrease
exponentially with growing $l$. In the weak coupling limit, Coulomb
interaction becomes less important in the low-energy regime, and
does not generate any anomalous dimension for $A$ and $v$. In this
case, $A(l)$ exhibits a complicated dependence on length scale $l$,
and $v(l)$ grows as certain powers of $l$. However, the residue
$Z_f(l)$ decreases as certain powers of $l$, thus the system
displays marginal FL behavior at low energies \cite{Isobe2016PRL}.

The aim of this work is to analyze the roles played by Coulomb
interaction and disorders based on the RG equations derived in both
the strong and weak coupling limits. Before studying the interplay
of Coulomb interaction and disorder, it is helpful to first drop
Coulomb interaction and only consider the fermion-disorder coupling,
which will be addressed in next section.

\section{Disorder effects in the absence of Coulomb interaction}
\label{Sec_clean_analysis}

In this section, we consider the influence of three types of
disorder on the low-energy properties of 2D SDFs by simply ignoring
the Coulomb interaction. We first make a detailed analysis of the
possible RG solutions by setting $\alpha = 0$, then discuss the
low-energy behaviors of renormalized parameters $A$ and $v$, and
finally compute the quasiparticle residue $Z_f$ to determine whether
disorders lead to non-FL behaviors of the system.

\subsection{Solutions of RG equations}\label{Sec_clean_solution}

In the non-interacting limit, the RG equations become
\begin{eqnarray}
\frac{d\ln A}{dl} &=& -C_g + z - 2,\label{Eq.RG_SA} \\
\frac{d\ln v}{dl} &=& -C_g + z - z_{1},\label{Eq.RG_Sv} \\
\frac{d\ln C_g}{dl} &=& 2C_D.\label{Eq.RG_SC_g}
\end{eqnarray}
Notice that the flow equation of $C_g$ is independent of the other
two equations, so it is possible to first obtain the solution of
$C_g$ and analyze its $l$-dependence.

In the case of random chemical potential, we have
\begin{eqnarray}
\frac{d C_g}{dl} &=& 2C_g^2,
\end{eqnarray}
which has the following solution
\begin{eqnarray}
C_g(l) = \frac{C_{g0}}{1 - 2C_{g0}l}.\label{Eq.cp_spec}
\end{eqnarray}
Here, $C_{g0}$ is the value of parameter $C_{g}(l)$ defined at the
upper energy limit $\Lambda$. There exists a characteristic length
scale $l_c = 1/2C_{g0}$. Below $l_c$, $C_{g}(l)$ is an increasing
function of $l$, thus random chemical potential is a marginal
relevant perturbation in the low-energy regime. As $l$ approaches
$l_c$ from below, $C_g$ appears to be divergent. We need to
understand such a superficial divergence with caution. The
perturbative RG calculations are reliable only for small values of
$C_g$, so we cannot expect to have a really divergent $C_g$.
Instead, the unbounded increase of $C_g$ with growing $l$ below
$l_c$ is most likely a signature that random chemical potential
causes an instability of the system and turns the semimetallic state
into a disorder controlled diffusive state \cite{Fradkin1986,
Shindou2009, Goswami2011PRL, Roy2014PRB, Syzranov2015,
Kobayashi2014, Biswas2014}.

For random mass, the equation for $C_g$ is
\begin{eqnarray}
\frac{d C_g}{dl} = -2C_g^2.
\end{eqnarray}
Its solution is
\begin{eqnarray}
C_g(l) = \frac{C_{g0}}{1 + 2C_{g0}l}.\label{Eq.mass_spec}
\end{eqnarray}
It is easy to verify that $C_g(l)\rightarrow 0$ as $l \rightarrow
+\infty$, which implies that random mass is a marginally irrelevant
perturbation to the system.

For random gauge potential, we find
\begin{eqnarray}
\frac{d C_g}{dl} = 0\Rightarrow C_g(l) = C_{g0}.
\label{Eq.gauge_spec}
\end{eqnarray}
It therefore turns out that random gauge potential is marginal and
$C_g$ is a $l$-independent constant.

We next use the flowing behavior of $C_g$ to analyze the impact of
disorders on the low-energy physical properties of 2D SDFs.
According to Ref.~\cite{Yang2014Nat.P}, the scaling parameters $z$
and $z_1$ can be fixed at $z_1 = z = 2$, thus both $A$ and $v$ are
taken to be marginal parameters at the starting point. We learn from
Eq.~(\ref{Eq.RG_SA}) and Eq.~(\ref{Eq.RG_Sv}) that $A$ and $v$ might
be substantially modified by disorders due to the existence of
$C_g$.

In the case of random chemical potential, $A$ and $v$ depend on $l$
as
\begin{eqnarray}
\frac{A(l)}{A_0} = \frac{v(l)}{v_0} = \sqrt{1 - 2C_{g0}l}.
\label{Eq.v_spec_chem}
\end{eqnarray}
As $l$ grows approaching $l_c$, both $A(l)$ and $v(l)$ vanish. For
random mass, we use Eq.~(\ref{Eq.mass_spec}) to obtain
\begin{eqnarray}
\frac{A(l)}{A_0} = \frac{v(l)}{v_0} =
\frac{1}{\sqrt{1+2C_{g0}l}},\label{Eq.v_spec}
\end{eqnarray}
which shows that $A(l)$ and $v(l)$ vanish in the limit $l
\rightarrow +\infty$. For random gauge potential, one can similarly
show that $A(l)$ and $v(l)$ decrease exponentially with growing $l$
and rapidly flow to zero as $l \rightarrow +\infty$. Therefore, we
conclude that all the three types of disorders can drive $A$ and $v$
to vanish in the lowest energy limit. Apparently, disorders have
completely different influences on $A$ and $v$ from Coulomb
interaction, which tends to increase $A$ and $v$ in the low-energy
regime.

\subsection{Quasiparticle residue \texorpdfstring{$Z_f$}{}}\label{Sec_Quasiparticle_Z}

In any interacting fermion system, the quasiparticle residue $Z_f$
serves as a crucial quantity to judge whether a correlated many
fermion system can be well described by normal FL theory: $Z_f$ is
finite in a normal FL, but vanishes in a marginal FL and a non-FL.
This quantity is usually defined by the wave renormalization
functions follows
\begin{eqnarray}
Z_f = \frac{1}{1 - \frac{\partial
\mathrm{Re}\Sigma^{R}(\omega)}{\partial\omega}}\label{Eq.def_Z},
\end{eqnarray}
where $\mathrm{Re}\Sigma^{R}(\omega)$ is the real part of retarded
fermion self-energy function. Taking advantage of the RG results, it
is more convenient to write it in the form \cite{WangLiu14}
\begin{eqnarray}
Z_f = e^{-\int_0^{l}C_gdl},
\end{eqnarray}
which explicitly includes the impact of disorder through the
parameter $C_g$. Making derivative with respect to $l$, we find that
\begin{eqnarray}
\frac{d\ln Z_f}{dl} = -C_g.
\end{eqnarray}
Therefore, the $l$-dependence of $Z_f(l)$ is indeed the same as that
of $A(l)$ and $v(l)$. For convenience, here we list the results as
follows:
\begin{eqnarray}
Z_f(l)&=& \left\{\begin{array}{ll}  \sqrt{1 - 2C_{g0}l}
&\textmd{Random chemical potential}
\\
\\
1\big/\sqrt{1+2C_{g0}l} &\textmd{Random mass}
\\
\\
e^{-C_{g0}l} &\textmd{Random gauge potential}
\end{array}\right.\nonumber \\ &&\label{Eq:ZfOnlyDisorderSummary}
\end{eqnarray}

For random chemical potential, $Z_f$ vanishes when $l$ approaches
$l_c$ from below. Recall that the fermion velocities also decrease
rapidly to zero at the same finite energy scale. These unusual
behaviors indicate the instability of Fermi liquid and may, as
aforementioned, signal the transition into a diffusive state.

For the other two types of disorders, $Z_f\rightarrow 0$ as
$l\rightarrow \infty$, indicating the breakdown of ordinary FL
description. However, there is a subtle difference in the
quantitative $l$-dependence of $Z_f$ between random mass and random
gauge potential. To illustrate this difference, we now compute the
fermion damping rate. For random mass, the corresponding $Z_{f}$ can
be approximated by
\begin{eqnarray}
Z_{f}(l)\propto\frac{1}{l}
\end{eqnarray}
for sufficiently large $l$. Using Eq.~(\ref{Eq.def_Z}) and the
scaling relation $\omega = \omega_{0}e^{-2l}$, where $\omega_0$ is
the upper limit of fermion energy, we obtain
\begin{eqnarray}
\mathrm{Re}\Sigma^{R}(\omega)\propto \omega
\ln\left(\frac{\omega_{0}}{\omega}\right).
\end{eqnarray}
Based on the Kramers-Kronig (KK) relation, we can compute the
imaginary part of retarded fermion self-energy and get
\begin{eqnarray}
\mathrm{Im}\Sigma^{R}(\omega)\propto\omega,
\end{eqnarray}
which is linear in $\omega$ and thus signals the appearance of
marginal FL behavior. For random gauge potential, we find that
\begin{eqnarray}
\textmd{Re}\Sigma^{R}(\omega) \propto \omega^{1-\delta/2},
\end{eqnarray}
where $\delta = C_{g0}$. Applying the KK relation leads us to
\begin{eqnarray}
\textmd{Im}\Sigma^{R}(\omega) \propto \omega^{1-\delta/2},
\end{eqnarray}
which implies that the system displays non-FL behavior since $C_{g0}
> 0$. We thus conclude that, random mass leads to marginal FL
behavior, whereas random gauge potential produces non-FL behavior.
As a comparison, random gauge potential \cite{Altland02} also
induces non-FL behavior of massless nodal fermions in $d$-wave
superconductors \cite{WangNJP2016}.

\section{Interplay of Coulomb interaction and disorder}
\label{Sec_Interplay_soultion}

In this section, we include both Coulomb interaction and disorder,
and investigate their mutual influence as well as the possible phase
transitions driven by their interplay. The consideration is based on
the full set of RG equations obtained in the strong and weak
coupling limits respectively. Following the procedure of
Sec.~\ref{Sec_clean_analysis}, we will determine the $l$-dependence
of $C_g$, specify the (ir)relevance of disorders, examine the
low-energy behaviors of $A$ and $v$, and finally study the physical
effects of Coulomb interaction.

\subsection{Random chemical potential}\label{Sec_R_che_p}

In this subsection, we present a detailed RG analysis in the case of
random chemical potential. The bare value of $\alpha$ is strongly
material dependent. For instance, $\alpha \approx 0.41$ in VO$_2$
\cite{Yang2010PRB, Pardo2009PRL, Banerjee2009PRL}, and $\alpha
\approx 0.44$ in black phosphorus \cite{Nagahama1985JPSJ,
Kim2015Sci}. The value of $\alpha$ could be greatly amplified when
the system is delicately tuned to certain quantum critical point.
Once $\alpha$ exceeds unity, i.e., $\alpha > 1$, the ordinary
perturbation expansion breaks down. However, since we perform RG
analysis by utilizing $1/N$ expansion, in principle $\alpha$ can
take a large value, and so does $\alpha_N$. As demonstrated in
Ref.~\cite{Isobe2016PRL}, the RG analysis is greatly simplified in
the large and small $\alpha_N$ limits, but the intermediate coupling
regime is technically hard to handle. In the following, we will
consider the strong and weak coupling limits separately. The same
procedure can be directly applied to study the cases of random mass
and random gauge potential, which are presented in the rest two
subsections.

\begin{figure*}[htbp]
\center\hspace{-0.90cm}
\subfigure{
\begin{minipage}{7.8cm}
 \includegraphics[width=2.7in]{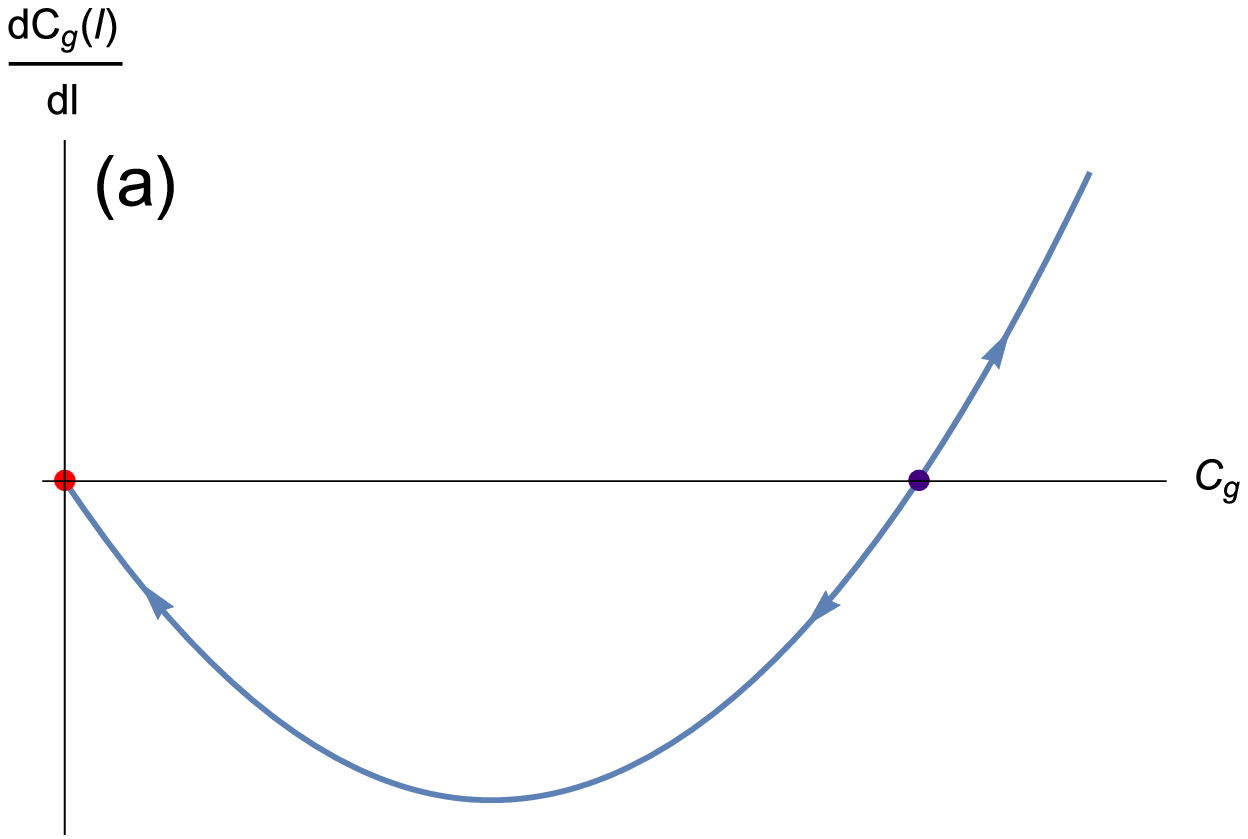}\label{Fig_Flow_dig_Chem}
 \vspace{0.5ex}\hspace{-0.20cm}
  \end{minipage}}
\subfigure{\begin{minipage}{7.8cm}
 \includegraphics[width=2.7in]{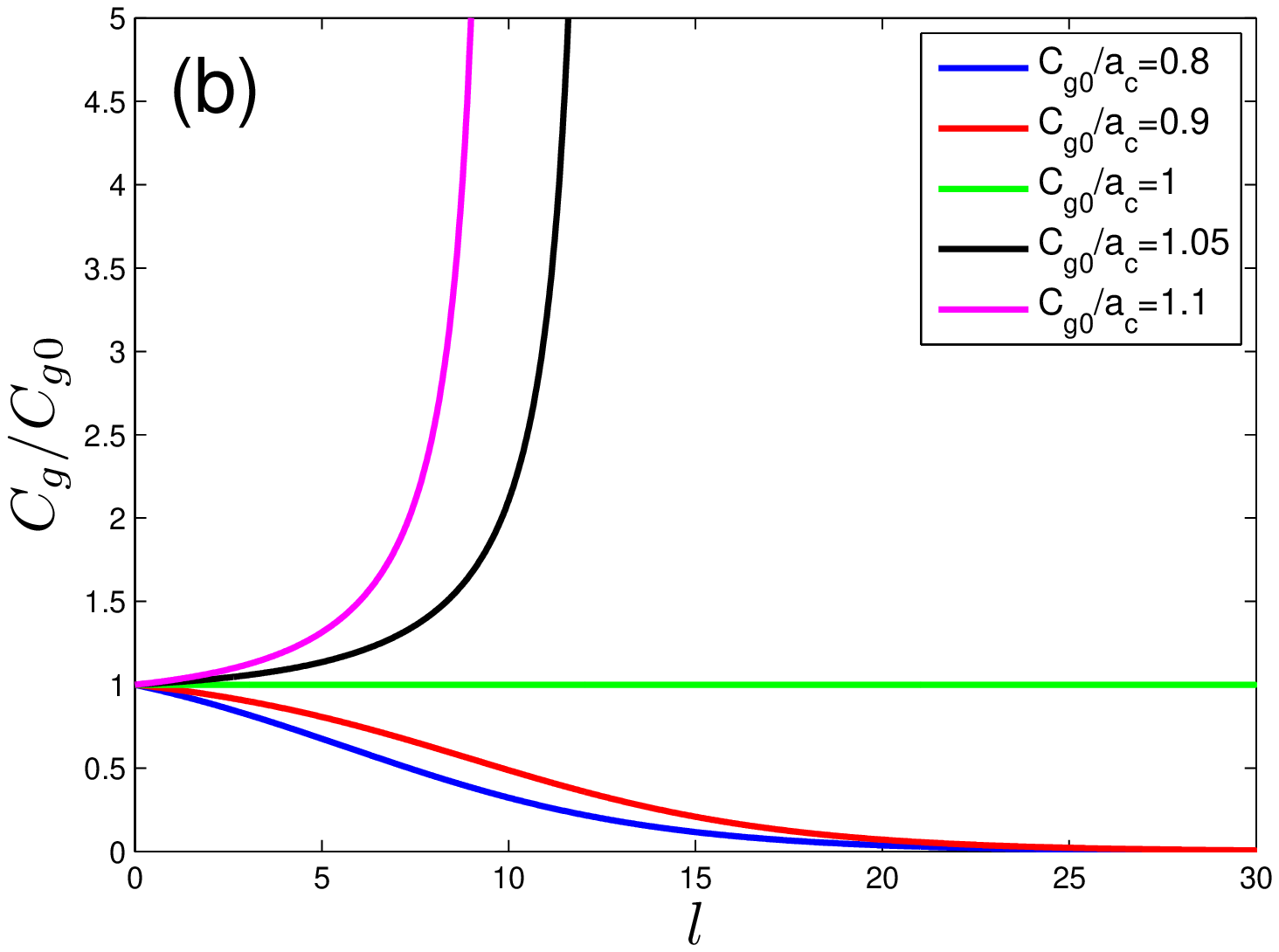}\label{Fig_CgStrongCP}
\end{minipage}}\vspace{0.1cm}
\subfigure{\begin{minipage}{7.8cm}
\includegraphics[width=2.55in]{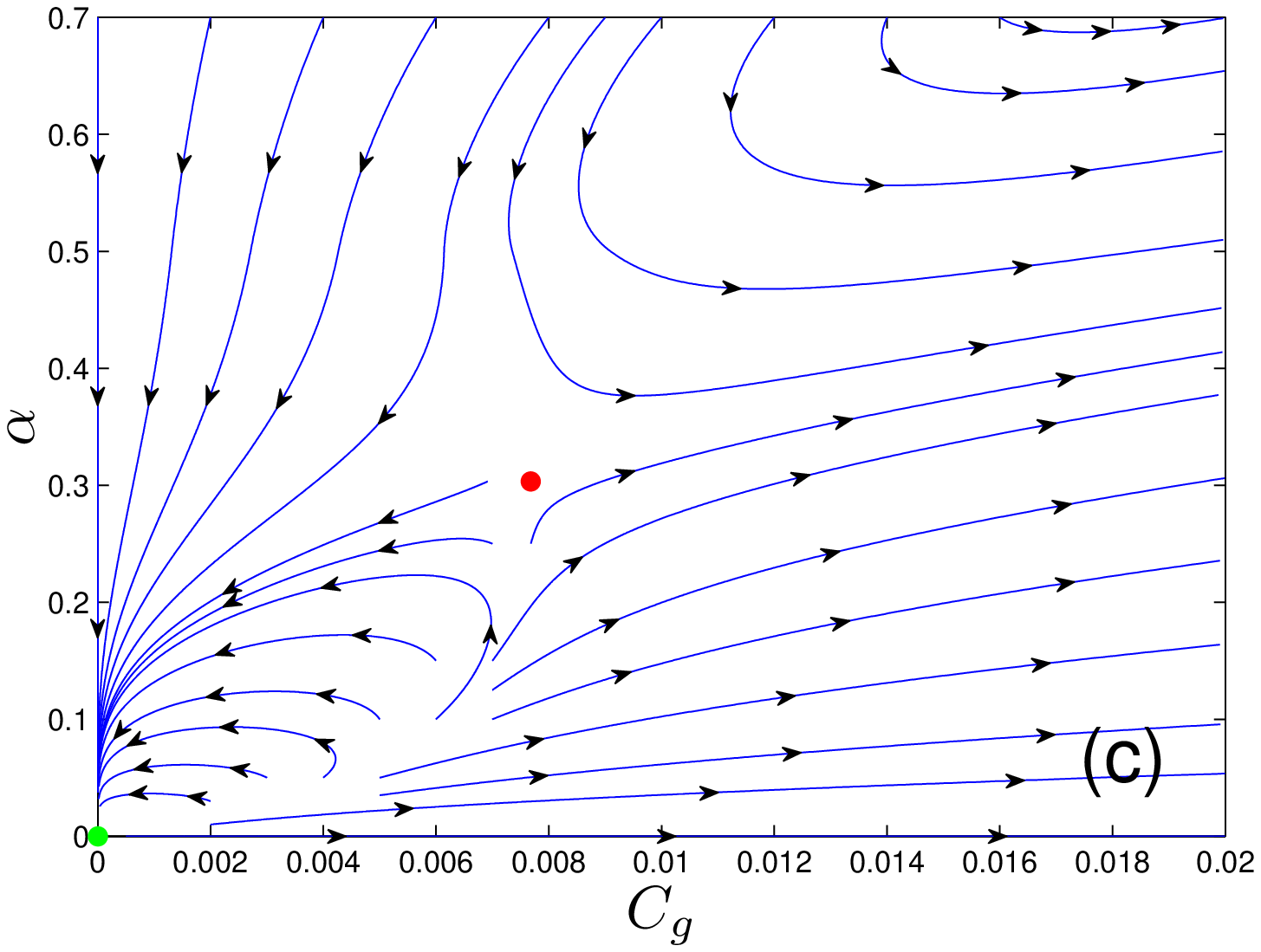}\label{Fig_WeakCouplingFlowCP}
\vspace{-0.05cm}
\end{minipage}\hspace{-0.25cm}}
\subfigure{\begin{minipage}{7.8cm}
\includegraphics[width=2.77in]{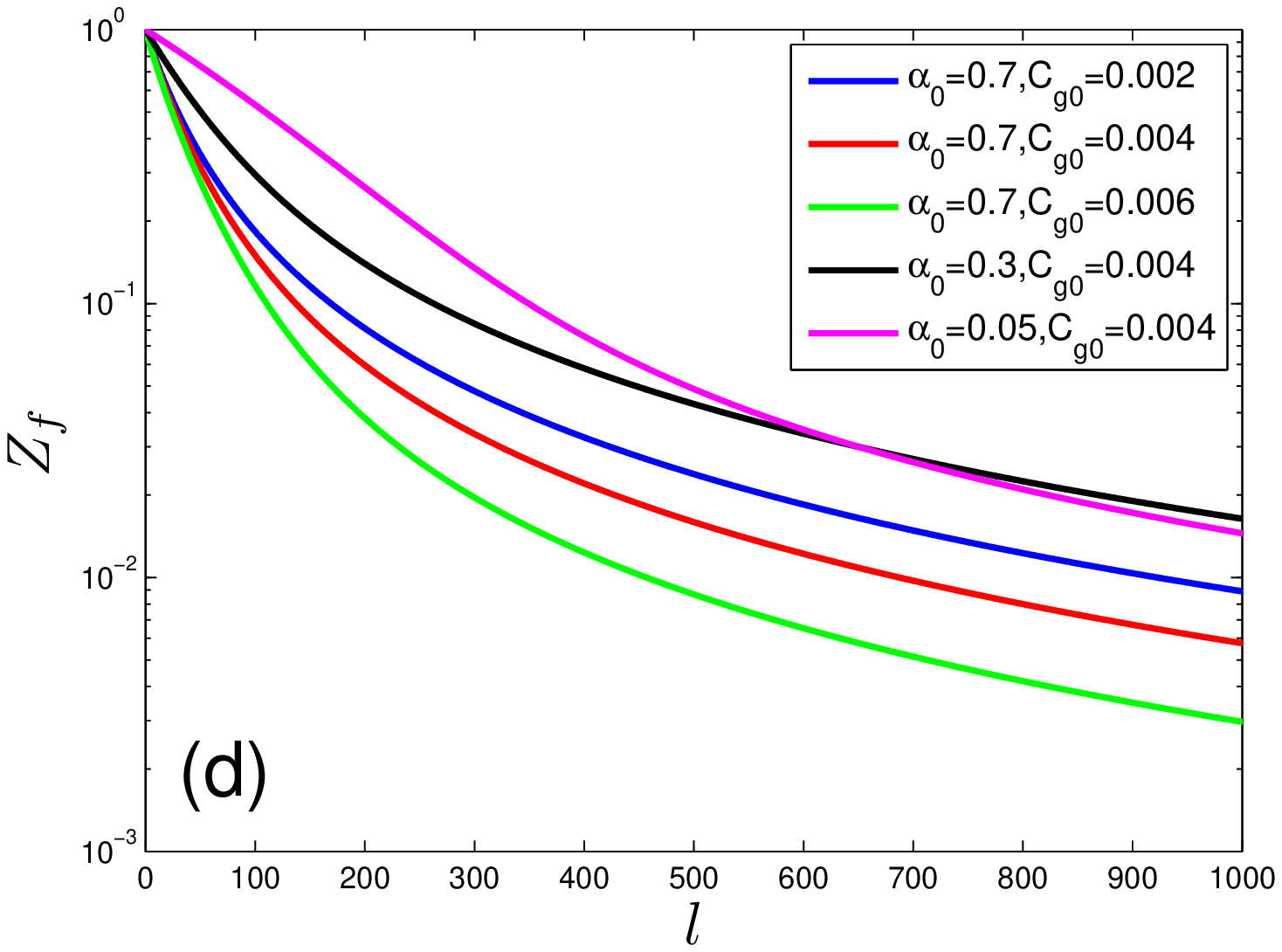}\label{Fig_Weak_Chem_Zf}\vspace{0.08cm}
\end{minipage}\hspace{0.30cm}}\vspace{-0.1cm}
\caption{(a) Flow diagram of $C_g$ in the case of random chemical
potential. There is a stable Gaussian fixed point $C_g = 0$ and a
finite unstable fixed point $C_g = a_c$. (b) Dependence of $C_{g}$
on the running scale for different initial values. (c) Flow diagram
in the parameter space spanned by $\alpha$ and $C_{g}$, with
$\alpha$ being in the weak coupling limit. Red point denotes the
stable Gaussian fixed point, and purple point represents the
unstable fixed point. (d) Flow of $Z_{f}$ obtained in the weak
coupling limit. Here, the fermion flavor is assumed to be $N =
2$.}\label{Fig_WeakRG}
\end{figure*}

\begin{figure*}[htbp]
\center \hspace{-6ex} \subfigure{
\includegraphics[width=2.7in]{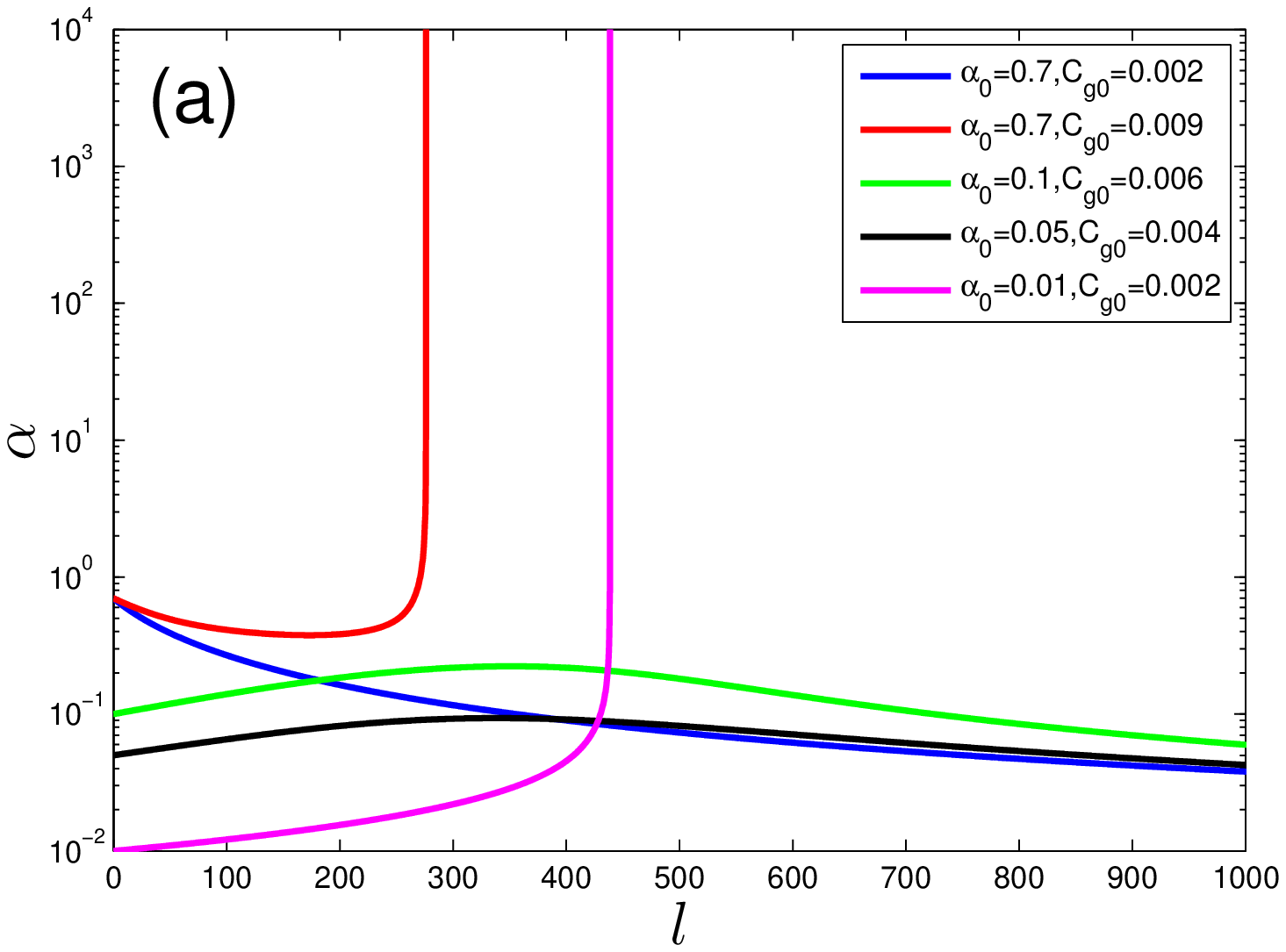}\label{Fig_WeakVRGCPAlpha}}
\hspace{3ex} \subfigure{
\includegraphics[width=2.7in]{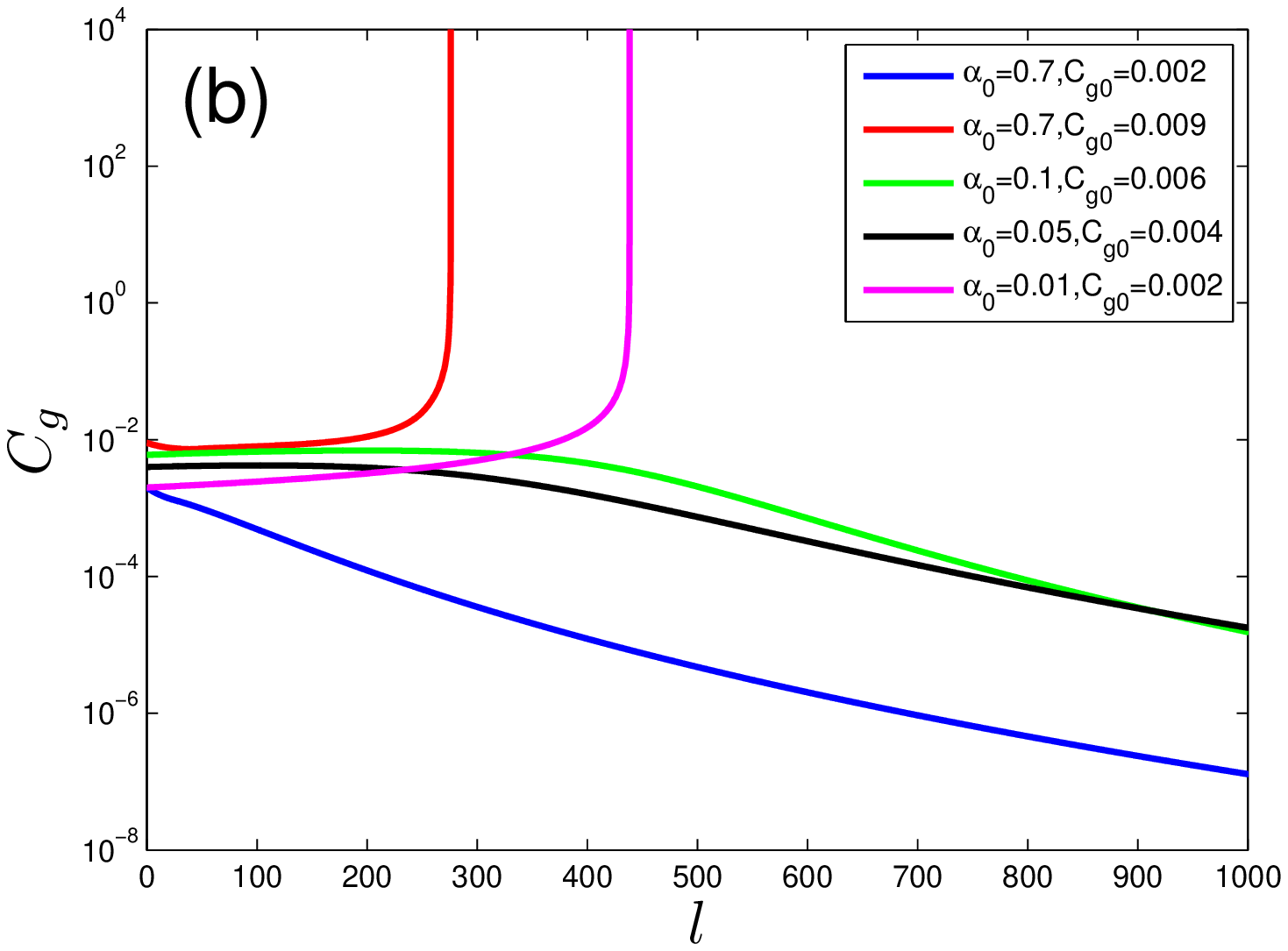}\label{Fig_WeakVRGCPCg}}\\
\hspace{-5ex}\subfigure{
\includegraphics[width=2.75in]{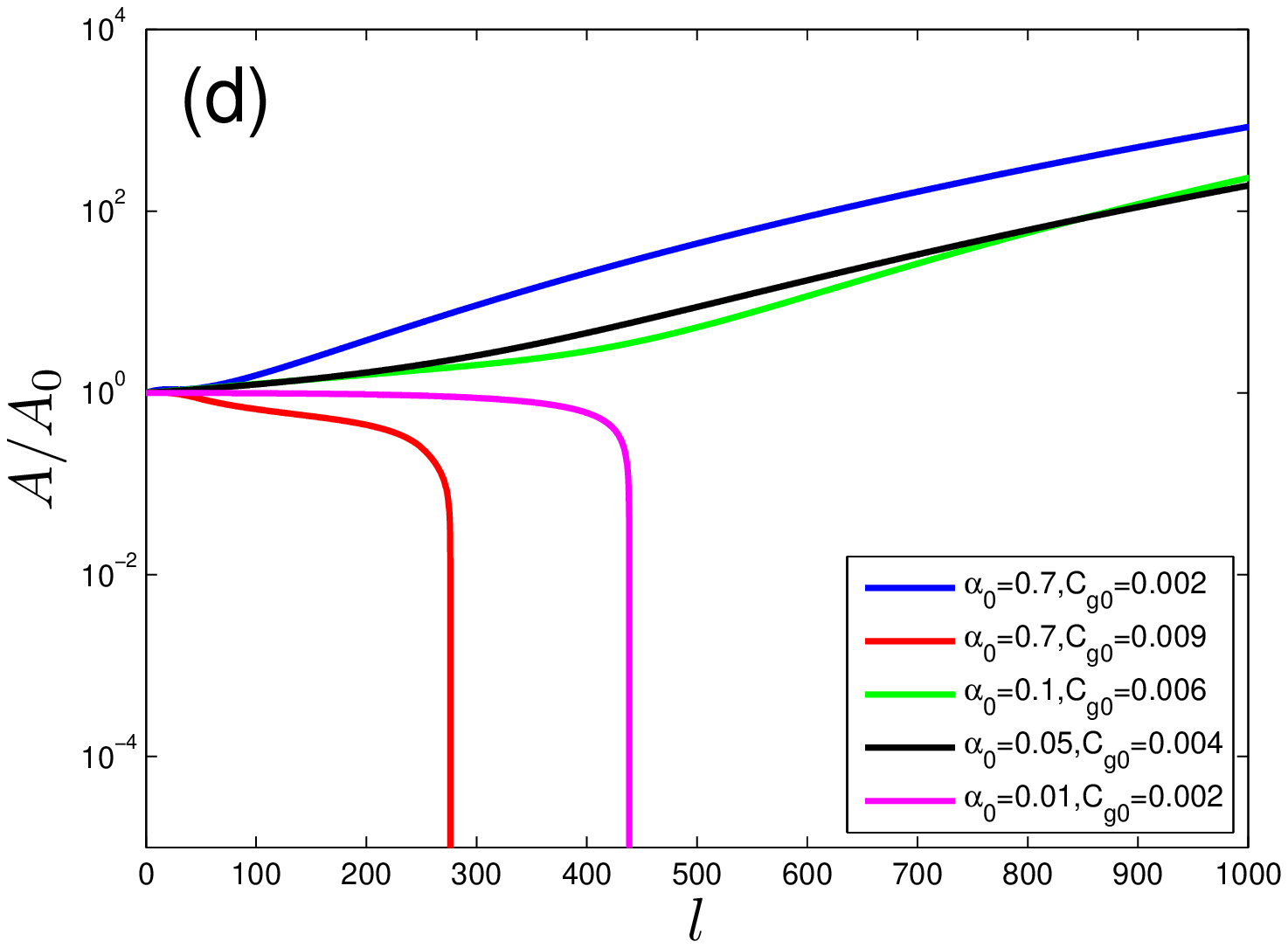}\label{Fig_WeakVRGCPA}}
\hspace{3ex} \subfigure{
\includegraphics[width=2.75in]{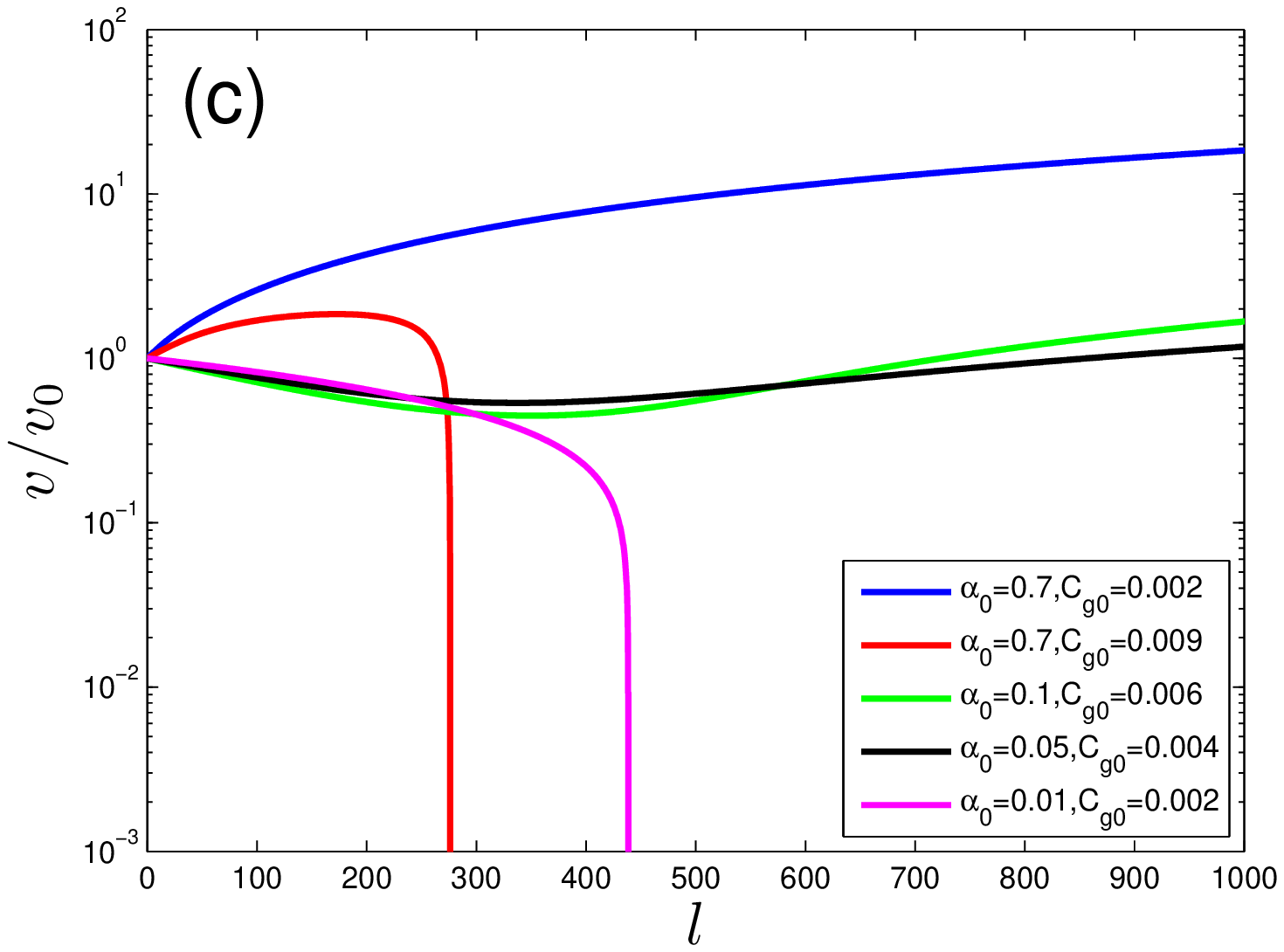}\label{Fig_WeakVRGCPVF}}\vspace{-0.1cm}
\caption{Flows of parameters $\alpha$, $C_{g}$, $A$, and $v$ are
shown in (a), (b), (c), and (d) respectively in the presence of random
chemical potential in the weak coupling limit. Here, the fermion
flavor is assumed to be $N = 2$.}\label{Fig_WeakRGPh}
\end{figure*}

\subsubsection{Strong coupling limit}\label{Sec_stong_c}

In the strong coupling limit, the flow equation of $C_g$ can be
written as
\begin{eqnarray}
\frac{dC_g}{dl} = (2 C_g - \gamma _A - \gamma _v)C_g,
\end{eqnarray}
which has the following solution
\begin{eqnarray}
C_g(l) = \frac{a_cC_{g0}}{C_{g0} - (C_{g0} - a_c)
e^{2a_cl}},\label{Eq_solu_C_g}
\end{eqnarray}
where
\begin{eqnarray}
a_c = \frac{\gamma_A + \gamma_v}{2} = \frac{0.2443}{N}\label{Eq_a_c}.
\end{eqnarray}

We have already illustrated in the last section that the parameter
$C_g$ for random chemical potential increases monotonously with
growing $l$ in the non-interacting limit with $\alpha_N = 0$. After
including the Coulomb interaction, there appears a critical value
$C_{g}^{\star} = a_c$. When $C_{g0} < a_{c}$, the parameter $C_{g}$
flows to zero in the long wavelength (low-energy) limit
$l\rightarrow \infty$, hence weak random chemical potential is a
irrelevant perturbation. Conversely, if $C_{g0} > a_{c}$, the
renormalized parameter $C_{g}$ increases rapidly with growing $l$
and formally flows to infinity at a finite length scale $l_c^{'} =
1/2a_c \ln \left[C_{g0}/(C_{g0} - a_c)\right]$. It thus turns out
that there are two infrared fixed points: $C_g^{\star} = 0$,
defining a stable Gaussian fixed point; $C_g^{\star} = a_c$,
corresponding to an unstable fixed point. The schematic flowing
behavior of $C_g$ is presented in
Figs.~\ref{Fig_Flow_dig_Chem}-~\ref{Fig_CgStrongCP}. As explained in
the last section, the most reasonable interpretation of the
unbounded increase of $C_g$ is that the system undergoes a quantum
phase transition, with the unstable fixed point $C_{g}^{\star} =
a_c$ being the quantum critical point. The Gaussian fixed point of
$C_g$ does not exist in the non-interacting limit, and is induced by
the addition of strong Coulomb interaction.

After identifying the fixed points of $C_g$, we turn to consider the
impact of random chemical potential on the fate of Coulomb
interaction strength $\alpha_N$. In a clean 2D SDF system, we know
from Ref.~\cite{Isobe2016PRL} that $\alpha_N$ decreases monotonously
as the energy scale is lowered. In the presence of weak random
chemical potential with $C_{g0} < a_{c}$, we infer from
Eq.~(\ref{Eq.RG_alpha}) that $\alpha_N \rightarrow 0$ in the
lowest-energy limit. In this case, random chemical potential is too
weak to induce any sizable change of the role played by the Coulomb
interaction. In contrast, when $C_{g0} > a_{c}$, $\alpha_N$
increases with growing $l$ and formally diverges as $l \rightarrow
l_c^{'}$. An apparent indication of this result is that the
importance of Coulomb interaction is significantly enhanced by
strong random chemical potential.

The superficial divergence of $\alpha_N$ and $C_{g}$ needs to be
properly understood. In the non-interacting limit with $\alpha_N =
0$, the unbounded increase of disorder parameter $C_g$ in the
low-energy regime could be interpreted as a quantum phase transition
of the system into a disorder-dominant diffusive state
\cite{Fradkin1986, Shindou2009, Goswami2011PRL, Roy2014PRB,
Syzranov2015, Kobayashi2014, Biswas2014}. However, if both $C_g$ and
$\alpha_N$ flow to very large values, the system may exhibit other
kinds of states. For instance, the system might be turned into an
excitonic insulator \cite{Stauber2005PRB, Vafek08}. Another
possibility is the formation of Wigner lattice, which is known to be
the ground state of 2D electron gas when the Coulomb interaction is
sufficiently strong \cite{Ceperley1989}. The nature of the ground
state of the system at large values $\alpha_N$ and $C_g$ is
currently unclear and needs to be further studied.

In the following, we focus on the low-energy properties of the
system at the Gaussian fixed point. Although $\alpha_N$ and $C_g$
both flow to zero in the low-energy regime at the Gaussian fixed
point, their physical effects on the system cannot be simply
neglected. Before eventually flowing to zero, Coulomb interaction
and random chemical potential lead to considerable corrections to
the quantities $A$ and $v$.

In the clean limit, both $A$ and $v$ increase as the energy is
lowered due to the Coulomb interaction. If the non-interacting SDFs
couple to random chemical potential, $A$ and $v$ are driven to
vanish at low energies. Apparently, Coulomb interaction and random
chemical potential give rise to distinct low-energy behaviors of $A$
and $v$. In the presence of both Coulomb interaction and random
chemical potential, we substitute Eq.~(\ref{Eq_solu_C_g}) into
Eq.~(\ref{Eq.RG_A}) and Eq.~(\ref{Eq.RG_v}) with chosen values $z_1
= z = 2$, and then obtain
\begin{eqnarray}
F(l) = F_0 e^{(\gamma_F - a_c)l}\sqrt{t_{c} -
(t_c-1)e^{2a_cl}},\label{Eq.Chem_v_A}
\end{eqnarray}
where $t_c \equiv C_{g0}/a_c$ and a function $F$ is introduced to
represent $A$ or $v$. As $C_g$ increases from the unstable fixed
point, where $t_c>1$, both $A$ and $v$ tend to vanish at a finite
length scale $l = l_c^{'}$. However, as $C_g$ flows to its Gaussian
fixed point, the function $F(l)$ exhibits the following asymptotic
behavior
\begin{eqnarray}
F(l)\big|_{l\rightarrow \infty}\sim F_0 e^{(\gamma_F-a_c)l}
\big|_{l\rightarrow \infty}.
\end{eqnarray}
Since $\gamma_A < a_c < \gamma_v$, we find that $A$ still flow to
zero, but $v$ grows indefinitely in the lowest energy limit.
Comparing to the clean case, the anomalous exponent $\gamma_A$
induced by Coulomb interaction is eliminated by random chemical
potential, whereas the anomalous exponent $\gamma_v$ is reduced.

The above results ought to be further examined. As $C_g$ approaches
the Gaussian fixed point, $\alpha_N$ flows to zero at very low
energies. However, this conclusion is obtained based on the
assumption that Coulomb interaction is in the strong coupling limit.
It is necessary to directly consider the case of weak Coulomb
interaction and verify whether the above results are reliable. This
will be presented in the next subsection.

\subsubsection{Weak coupling limit}\label{Sec_weak_chem_lim}

\begin{figure*}[htbp]
\center\hspace{-0.9cm}\vspace{-0.1cm}
\subfigure{
\begin{minipage}{7.8cm}
\includegraphics[width=2.7in]{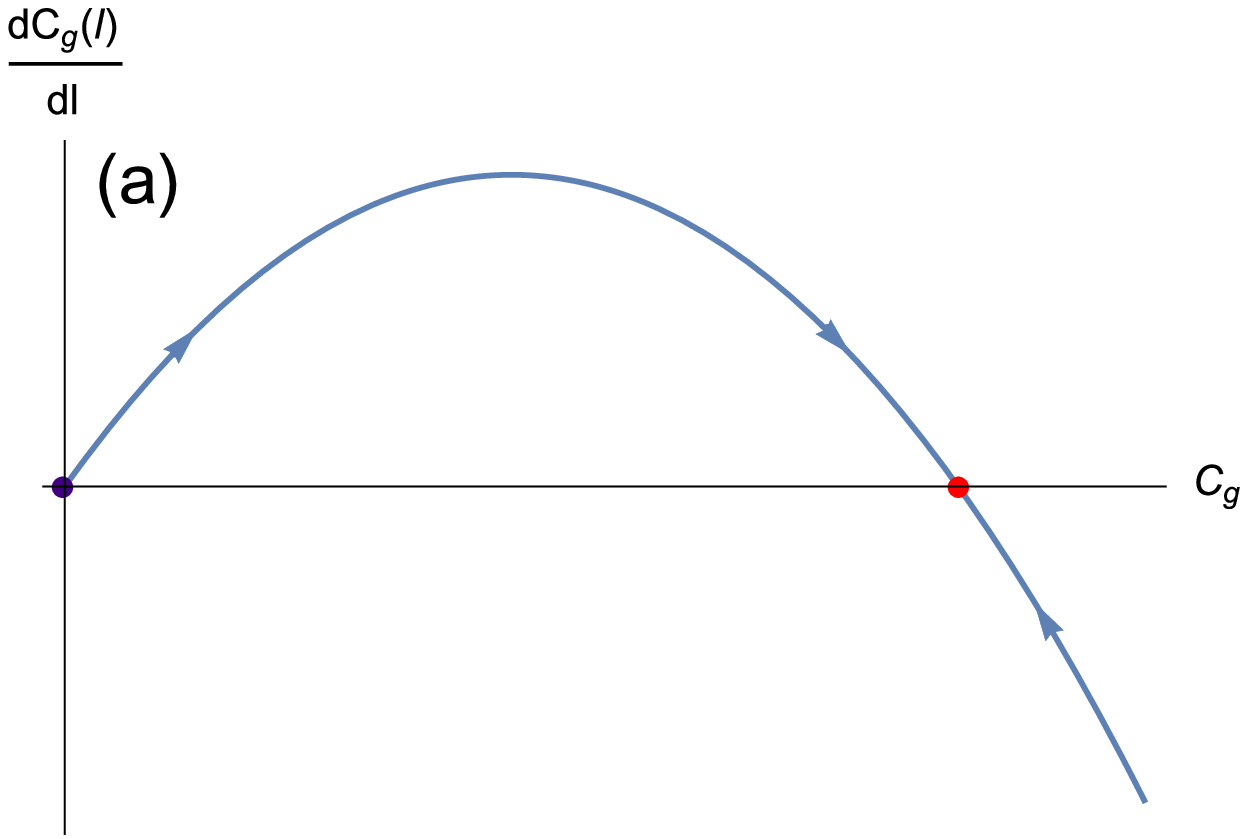}\label{Fig_WeakVRGMass}\vspace{0.1cm}
\end{minipage}\hspace{-0.1cm}}
\subfigure{\begin{minipage}{7.8cm}
\includegraphics[width=2.68in]{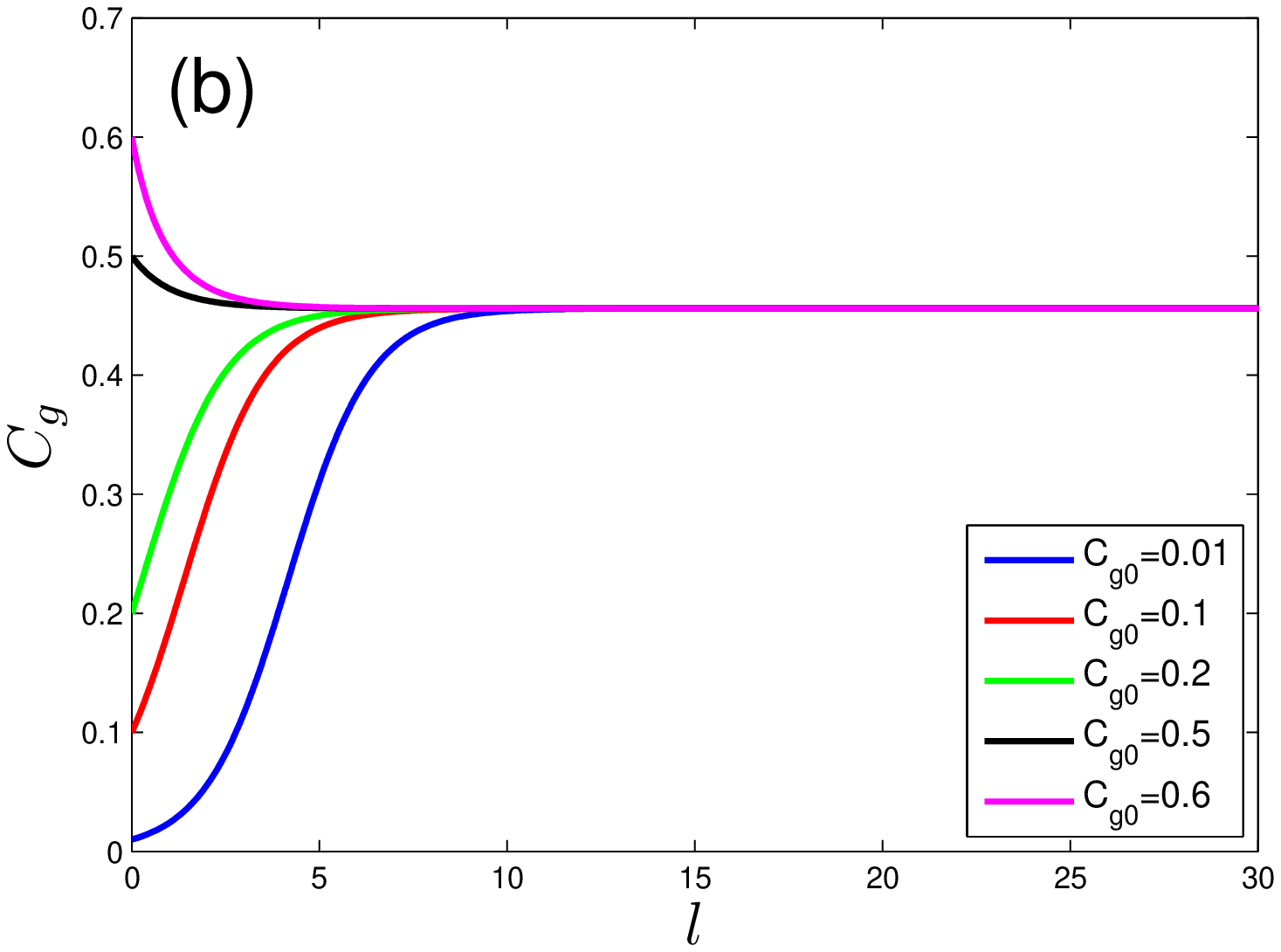}\label{Fig_CgStrongMass}
\end{minipage}}\vspace{0.1cm}
\subfigure{\begin{minipage}{7.8cm}
\includegraphics[width=2.55in]{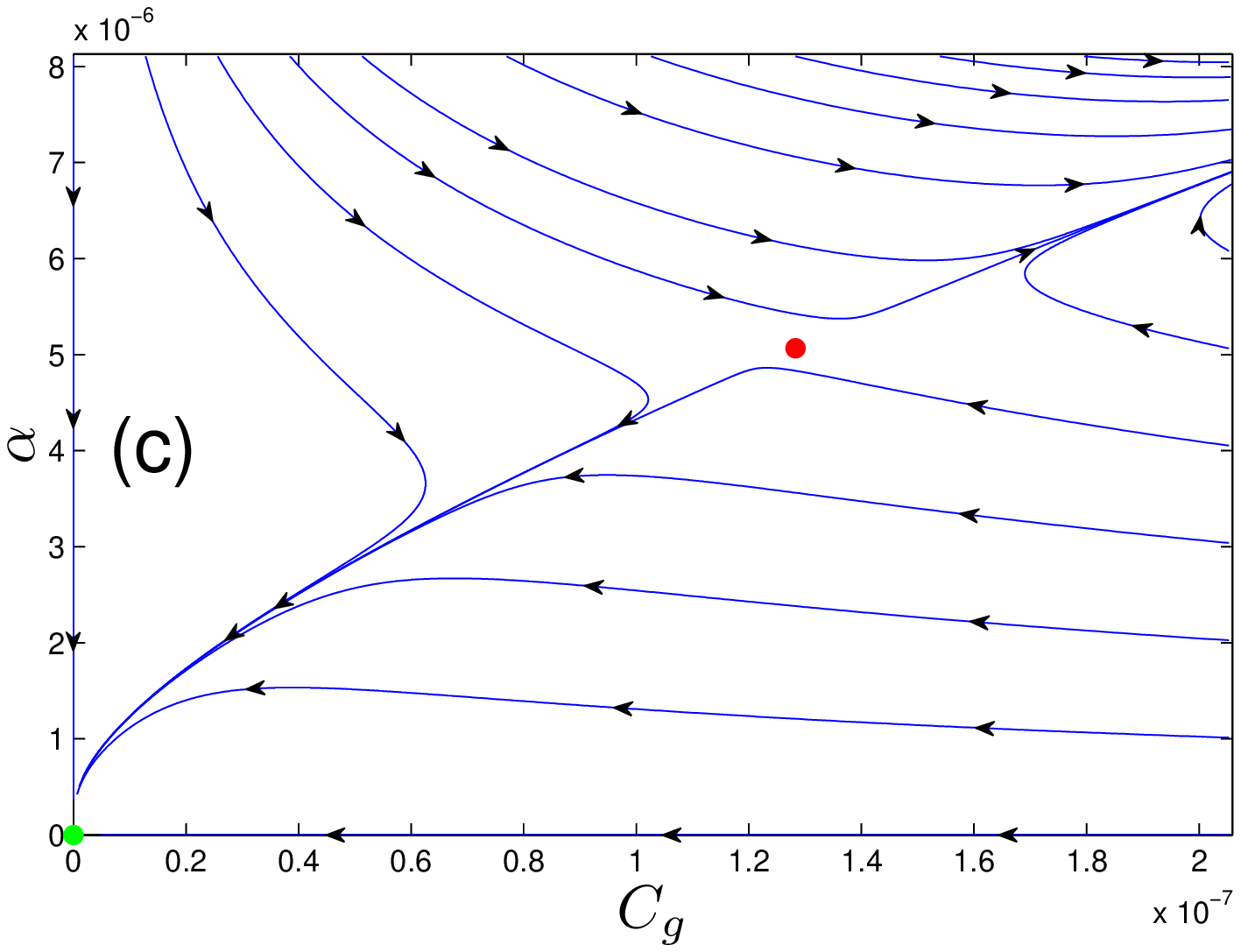}\label{Fig_WeakCouplingFlowMass}
\end{minipage}\hspace{-0.40cm}}
\subfigure{\begin{minipage}{7.8cm}
\includegraphics[width=2.85in,height=2.03in]{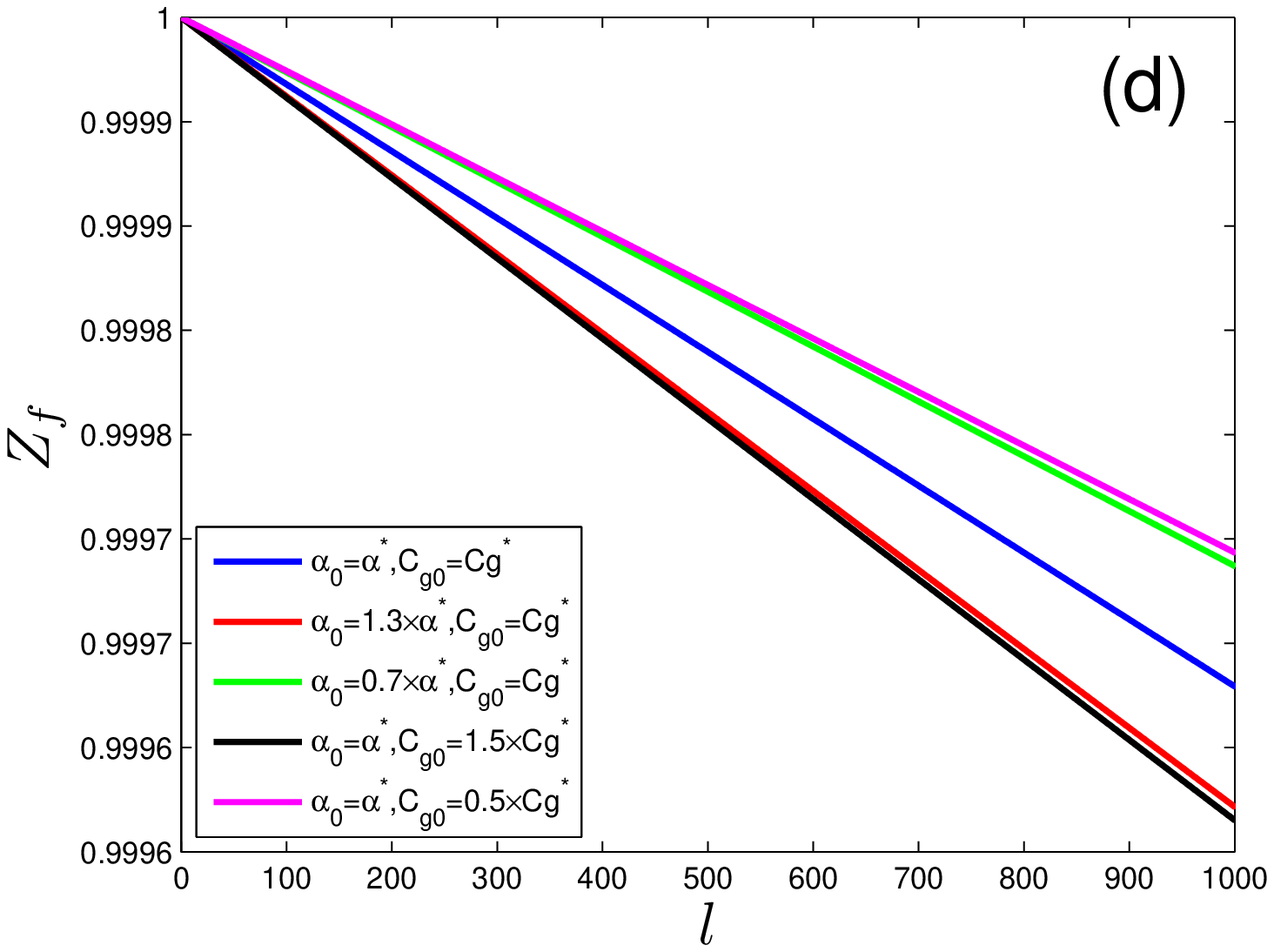}\label{Fig_WeakVRGMassZf}
\end{minipage}\hspace{0.61cm}}\vspace{-0.1cm}
\caption{(a) Flow diagram of $C_g$ for random mass obtained in the
strong coupling limit. There is a table fixed point $C_g = a_m$. (b)
Dependence of $C_{g}$ on running length scale $l$ for different
initial values of $C_{g0}$. The existence of stable fixed point of
$C_g$ can be clearly seen. (c) Flow diagram in the space spanned by
$\alpha$ and $C_{g}$ for random mass, where $\alpha$ is in the weak
coupling limit. The purple and red points
represent two fixed points. (d) Flow of $Z_{f}$
due to interplay of weak Coulomb interaction and random mass. Here,
the fermion flavor is assumed to be $N = 2$.}\label{Fig:WeakVRGMass}
\end{figure*}
We now consider the case of weak coupling limit with $\alpha_N \ll
1$. In the case of random chemical potential, we obtain the flow
equations:
\begin{eqnarray}
\frac{d\ln \alpha_N}{dl} &=& C_g -
\frac{\alpha_N}{4\pi^2N},\label{Eq.RG_alpha_weak}\\
\frac{d\ln C_g}{dl} &=& 2C_g -\frac{\alpha_N|\ln\alpha_N|}{2\pi^2N}
- \frac{\alpha_N}{4\pi^2N}.\label{Eq.RG_Cg_weak}
\end{eqnarray}
By demanding these two equations to vanish, it is easy to find two
infrared fixed points, namely a Gaussian fixed point
$(\alpha_N^{\star}, C_g^{\star}) = (0,0)$ and a finite fixed point
\begin{eqnarray}
(\alpha_N^{\star}, C_g^{\star}) = \left(1/\sqrt{e},1/4\pi^2N
\sqrt{e}\right).
\end{eqnarray}
The concrete flow diagram of $\alpha$ and $C_{g}$ is plotted in
Fig.~\ref{Fig_WeakCouplingFlowCP}, which shows that the finite fixed
point is unstable. Upon leaving this point, $\alpha_N$ and $C_g$
either flow to the Gaussian fixed point, or flow to larger values
with growing $l$. Unfortunately, it is not clear how $\alpha_N$ and
$C_g$ evolute to the strong coupling regimes due to our poor
knowledge of the intermediate regime of Coulomb interaction. In the
following, we concentrate on the low-energy properties of the system
in the vicinity of the Gaussian fixed point. The flows of $C_{g}$,
$\alpha_N$, $A$, and $v$ with varying $l$ are depicted in
Figs.~\ref{Fig_WeakVRGCPAlpha}-\ref{Fig_WeakVRGCPVF}, respectively.
It is interesting that $v$ exhibits a non-monotonic dependence on
$l$ for some specific initial values. When $\alpha_N$ and $C_g$ flow
to the Gaussian fixed point, $A(l)$ and $v(l)$ display nearly linear
dependence on $l$ for large $l$, which implies that $A(l)$ and
 $v(l)$ still receive weak logarithmic-like corrections before
$\alpha_N$ and $C_g$ flowing to zero.

\subsubsection{Quasiparticle residue \texorpdfstring{$Z_f$}{}}\label{Sec_Z_chem}

After specifying the possible fixed points, we now examine whether
the system exhibits non-FL behaviors. Now the quasiparticle residue
$Z_f$ will be recalculated after taking into account the interplay
of Coulomb interaction and random chemical potential. Using
Eq.~(\ref{Eq.def_Z}), we have
\begin{eqnarray}
Z_f = e^{-\int_0^{l}(\gamma_z + C_g)dl}.\label{Eq.Z_full}
\end{eqnarray}
Although $\alpha_N$ and $C_g$ flow to the trivial fixed point in the
weak coupling limit, $A(l)$ and $v(l)$ receive logarithmic
corrections. The $l$-dependence of $Z_{f}$ is shown in
Fig.~\ref{Fig_Weak_Chem_Zf}, which manifests that $Z_{f}(l)$
decreases much more slowly than an exponential function. In
particular, in the weak coupling limit, random chemical potential
can be nearly ignored and weak Coulomb interaction governs the
low-energy behaviors of the system, which exhibits non- and marginal
FL behaviors over a wide range of energy regimes
\cite{Isobe2016PRL}.

\subsection{Random mass}\label{Sec_Ran_mass}

\begin{figure*}[htbp]
\center \hspace{-6.5ex} \subfigure{
\includegraphics[width=2.7in]{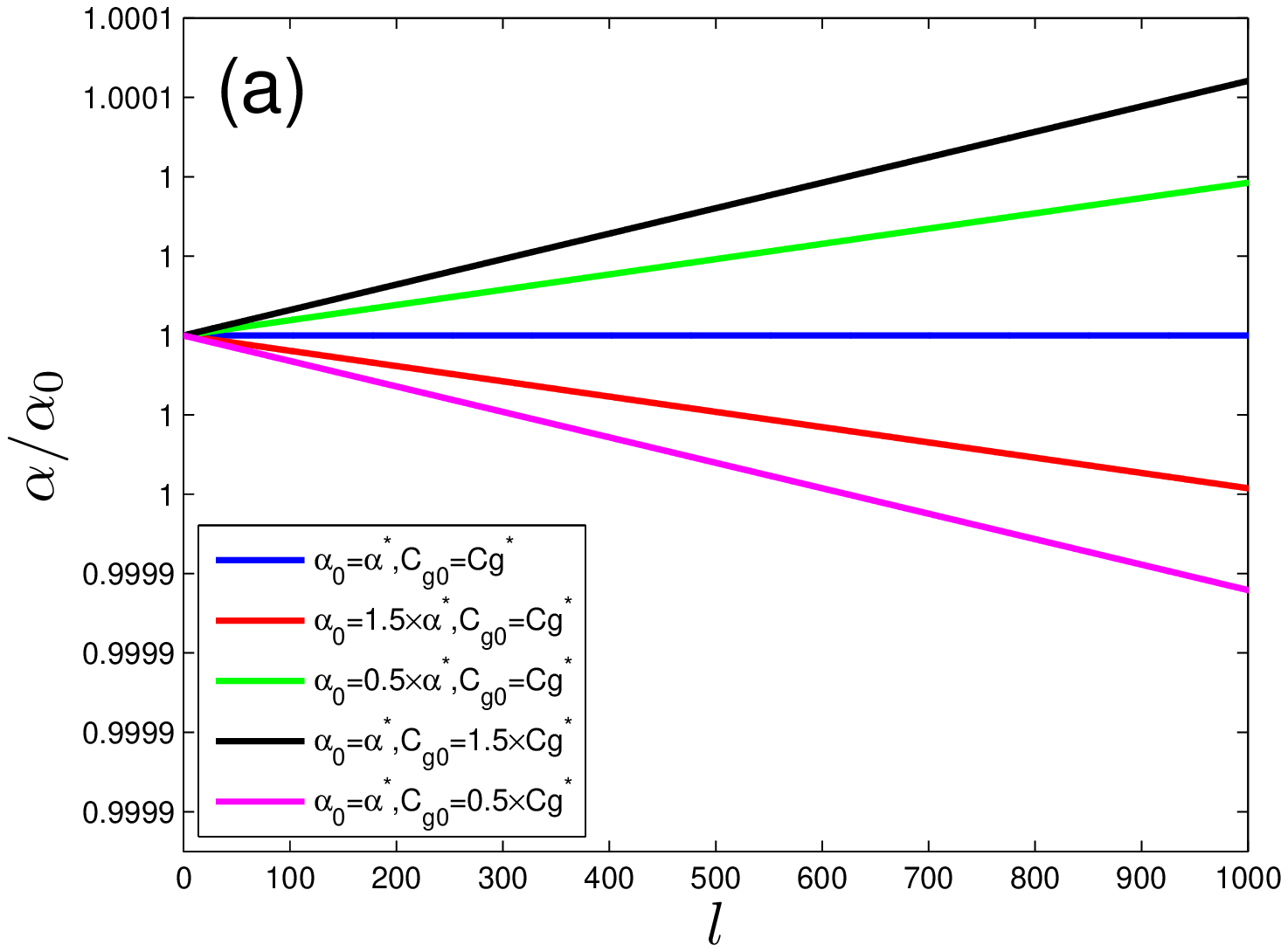}\label{Fig_WeakVRGMassAlpha}}
\hspace{1.5ex} \subfigure{
\includegraphics[width=2.75in]{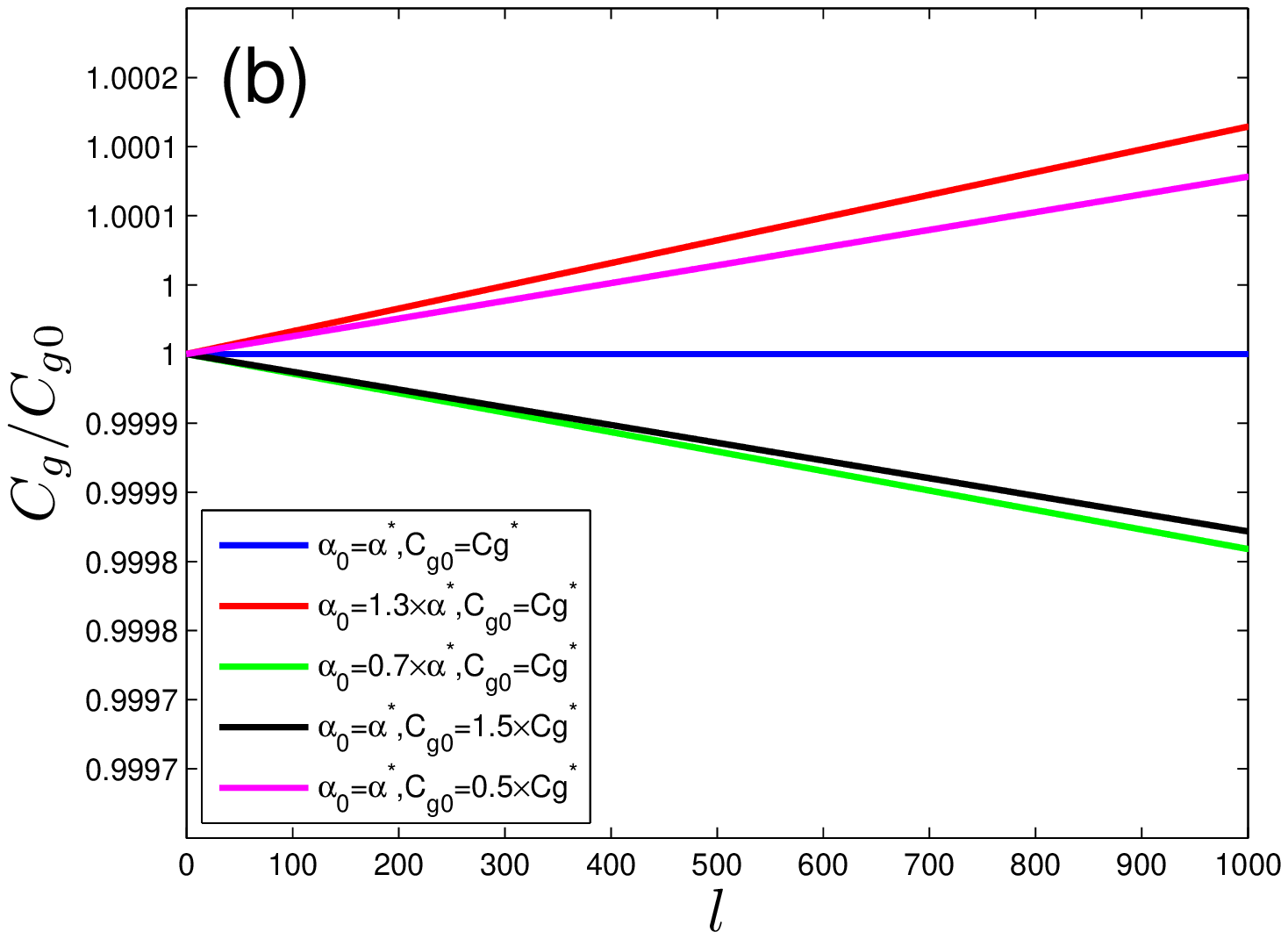}\label{Fig_WeakVRGMassCg}}\\
\hspace{-6.5ex} \subfigure{
\includegraphics[width=2.68in]{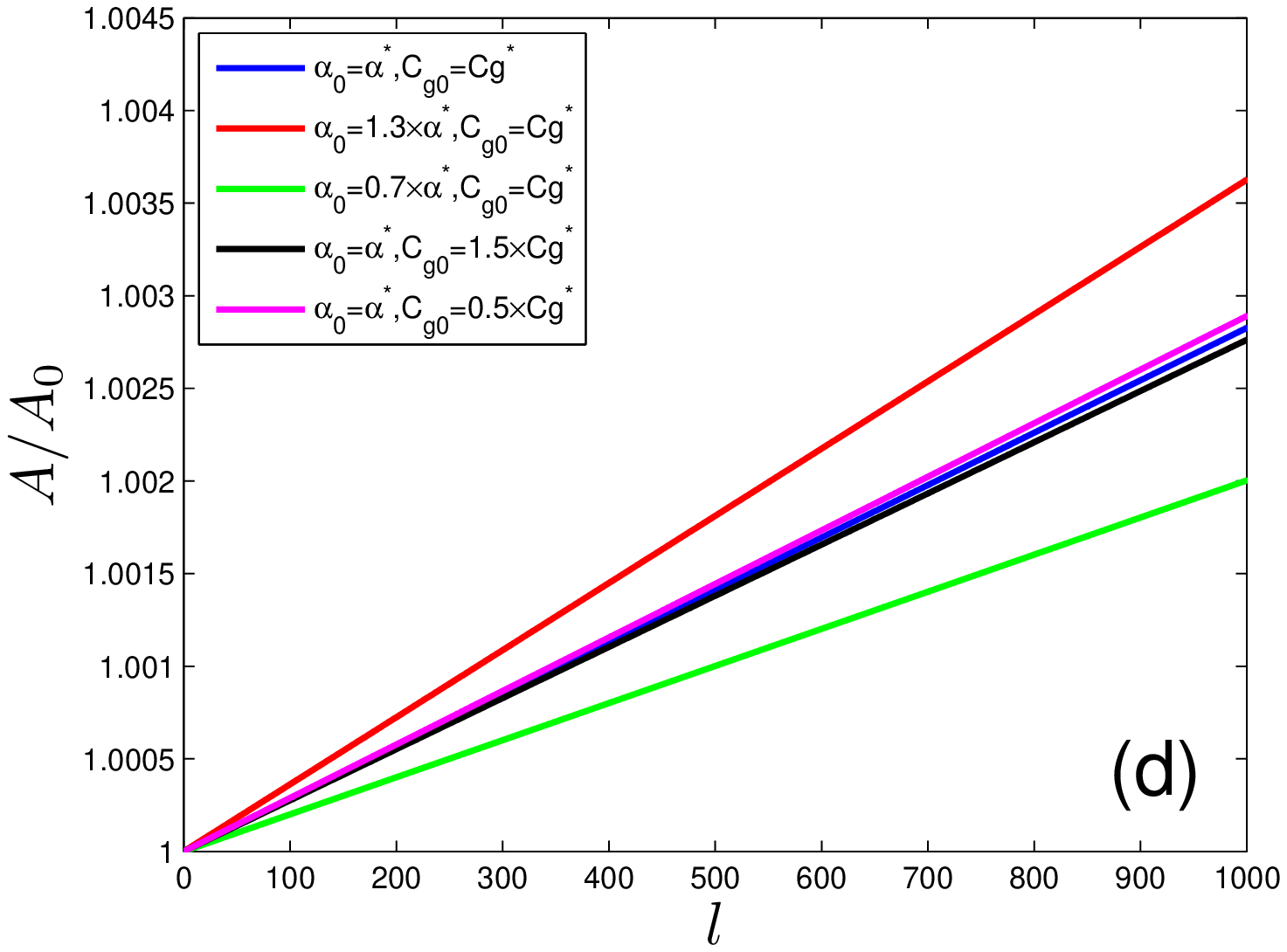}\label{Fig_WeakVRGMassA}}
\hspace{2ex} \subfigure{
\includegraphics[width=2.68in]{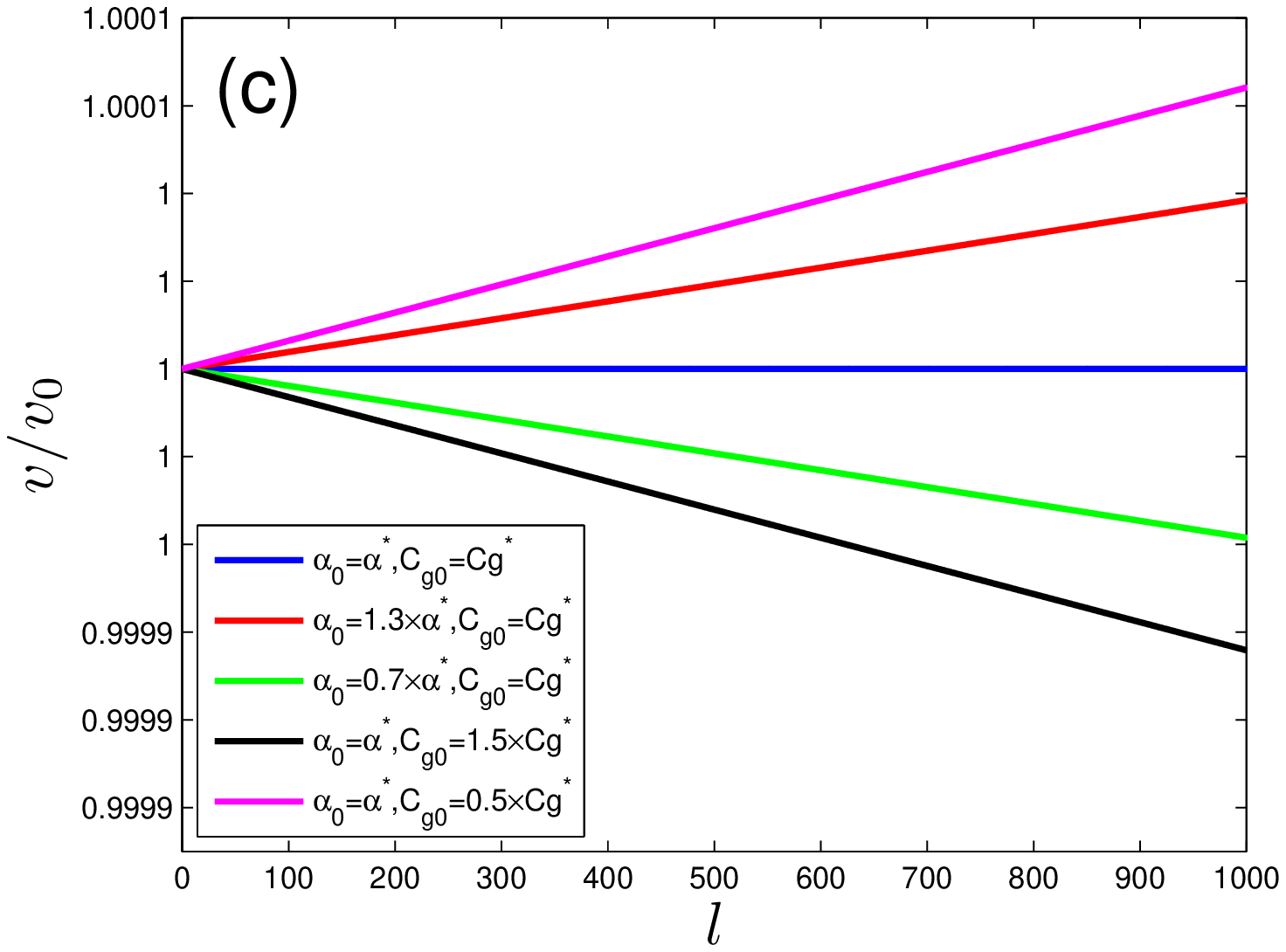}\label{Fig_WeakVRGMassVF}}\vspace{-0.1cm}
\caption{Flows of $\alpha$, $C_{g}$, $A$, and $v$ due to interplay
of Coulomb interaction and random mass are shown in (a), (b), (c),
and (d) respectively. The results are obtained in the weak coupling
limit of $\alpha$. $\alpha^{\star}$ and $C_g^{\star}$ stand for the
values of $\alpha$ and $C_g$ at the unstable fixed point,
respectively. Here, the fermion flavor is assumed to be $N=2$.}
\label{Fig:WeakVRGMassPh}
\end{figure*}
In this subsection, we consider the interplay of Coulomb interaction
and random mass, and show that it results in quite different
properties in the strong coupling limit comparing to the case of
random chemical potential. However, in the weak coupling limit, the
RG flow for the strength of random mass is similar, but not
identical, to random chemical potential.

\subsubsection{Strong Coupling limit}

In the strong coupling regime with $\alpha_{N} \gg 1$, the RG equation
of $C_{g}$ is given by
\begin{eqnarray}
\frac{dC_g}{dl} = (-2C_g + 2\gamma_{D} - \gamma _A - \gamma_v)C_g,
\end{eqnarray}
where $\gamma_D = \frac{1.1564}{N}$. Its solution has the following
form
\begin{eqnarray}
C_g(l) = \frac{a_mC_{g0} e^{2a_ml}}{C_{g0}(e^{2a_ml} - 1) +
a_m},\label{Eq_solu_C_g_mass}
\end{eqnarray}
where
\begin{eqnarray}
a_m = \gamma_{D} - \frac{\gamma_A + \gamma_v}{2} = \frac{0.9121}{N}.
\end{eqnarray}
The parameter $C_g$ has two infrared fixed points: an unstable fixed
point $C_g^{\star} = 0$, and a stable fixed point $C_g^{\star} = a_m$.
The existence of these two fixed points can be clearly seen from
Figs.~\ref{Fig_WeakVRGMass} - \ref{Fig_CgStrongMass}. Recall that
random mass is marginally irrelevant in the non-interacting limit.
Our RG analysis show that the strong Coulomb interaction promotes
random mass to become marginally relevant. In addition, it can be
deduced from Eq.~(\ref{Eq.RG_alpha}) that the Coulomb interaction
parameter $\alpha$ increases indefinitely as $l \rightarrow
+\infty$. Comparing this result to the clean limit
\cite{Isobe2016PRL}, we can see that the importance of Coulomb
interaction is also significantly enhanced by random mass provided
that the initial value of $\alpha_N$ is sufficiently large.
Therefore, the roles played by strong Coulomb interaction and random
mass are both substantially promoted by their interplay.

To illustrate the mutual promotion, we now analyze the low-energy
behaviors of parameterd $A$ and $v$, and discuss how they are
influenced by the interplay of Coulomb interaction and random mass.
Substituting Eq.~(\ref{Eq_solu_C_g_mass}) into Eq.~(\ref{Eq.RG_A})
and Eq.~(\ref{Eq.RG_v}), then solving the differential
equations, we obtain
\begin{eqnarray}
F(l) = \frac{F_0 e^{\gamma_F l}}{\sqrt{t_{m}(e^{2a_ml}-1) + 1}},
\end{eqnarray}
where $t_m \equiv C_{g0}/a_m$ and once again $F$ stands for $A$ or
$v$. Since $0 < \gamma_F < a_m$, for very large $l$, $F(l)$ behaves
asymptotically as
\begin{eqnarray}
F(l)\big|_{l\rightarrow \infty}\sim F_0 e^{(\gamma_F-a_m)l}
\big|_{l\rightarrow \infty}\rightarrow 0,
\end{eqnarray}
which shows that $A(l)$ and $v(l)$ both vanish at large $l$.

Similar to the case of random chemical potential, the parameter
$\alpha_N$ also flows to very large values without upper bound, but
the parameter $C_g$ for random mass flows to certain finite value.
Based on Eq.~(\ref{Eq.Z_full}), $Z_f$ can be written as
\begin{eqnarray}
Z_f(l)\big|_{l \rightarrow +\infty} = e^{-\delta l},
\label{Eq.Z_f_rm}
\end{eqnarray}
where $\delta > a_m$ for large $\alpha_N$. Making the same analysis
as presented in Sec.~\ref{Sec_Quasiparticle_Z}, we find that the
interplay of strong Coulomb interaction and random mass drives the
system to become a non-FL, even in the lowest energy limit.

\subsubsection{Weak coupling limit}

We now study the interplay of Coulomb interaction and random mass
supposing the initial value of strength of Coulomb interaction is
small. Following the computational steps presented in
Sec.~\ref{Sec_weak_chem_lim}, we have solved the corresponding RG
equations. According to the flow diagram depicted in
Fig.~\ref{Fig_WeakCouplingFlowMass}, there are a stable fixed point
$(\alpha_N^{\star}, C_g^{\star}) = (0,0)$ and an unstable fixed
point
\begin{eqnarray}
(\alpha_N^{\star}, C_g^{\star}) = \left(e^{-\frac{23}{2}},
e^{-\frac{23}{2}}/4\pi^2N \right).
\end{eqnarray}
The detailed $l$-dependence of the parameters $\alpha_N$, $C_{g}$,
$A$, and $v$ are presented in
Figs.~\ref{Fig_WeakVRGMassAlpha} - \ref{Fig_WeakVRGMassVF},
respectively. From Fig.~\ref{Fig_WeakVRGMassAlpha} and
Fig.~\ref{Fig_WeakVRGMassCg}, we observe that $\alpha$ and $C_{g}$
flow to distinct regimes when their initial values take different
values. Moreover, Fig.~\ref{Fig_WeakVRGMassA} and
Fig.~\ref{Fig_WeakVRGMassVF} show that, $A(l)$ always increases over
a broad range of $l$, but $v(l)$ may either increase or decrease,
depending on the initial values of $\alpha_N$ and $C_{g}$.

From the above analysis made in the strong and weak coupling regimes
of Coulomb interaction, we conclude that the low-energy behaviors of
2D SDFs are mainly determined by the initial value of interaction
parameter $\alpha_N$. The SDF system exhibits distinct properties at
large and small values of $\alpha_N$. The quasiparticle residue
$Z_f$ due to strong Coulomb interaction and random mass is already
given in Eq.~(\ref{Eq.Z_f_rm}). In the weak coupling limit, both
$\alpha_N$ and $C_g$ flow to zero. In this case, $Z_f(l)$ cannot be
expressed in such a power-law function as Eq.~(\ref{Eq.Z_f_rm}). We
plot the variation of $Z_{f}$ with $l$ in
Fig.~\ref{Fig_WeakVRGMassZf}, which exhibits that $Z_{f}(l)$
decreases much more slowly than an exponential function. Thus, in
the weak coupling limit, random mass does not induce remarkably
qualitative change for the behavior of $Z_{f}$ if the system is
driven to the Gaussian fixed point $(\alpha_N^{\star}, C_g^{\star})
= (0,0)$. As $\alpha_N$ decreases from certain large value down to
some small value, the SDF system is altered from a non-FL to a
marginal FL. However, since the intermediate coupling regime is
technically hard to access, it remains unclear whether this is a
phase transition or a crossover, and further research is required to
address this issue.

\begin{figure*}[htbp]
\center \hspace{-3.0ex}\subfigure{
\includegraphics[width=2.6in]{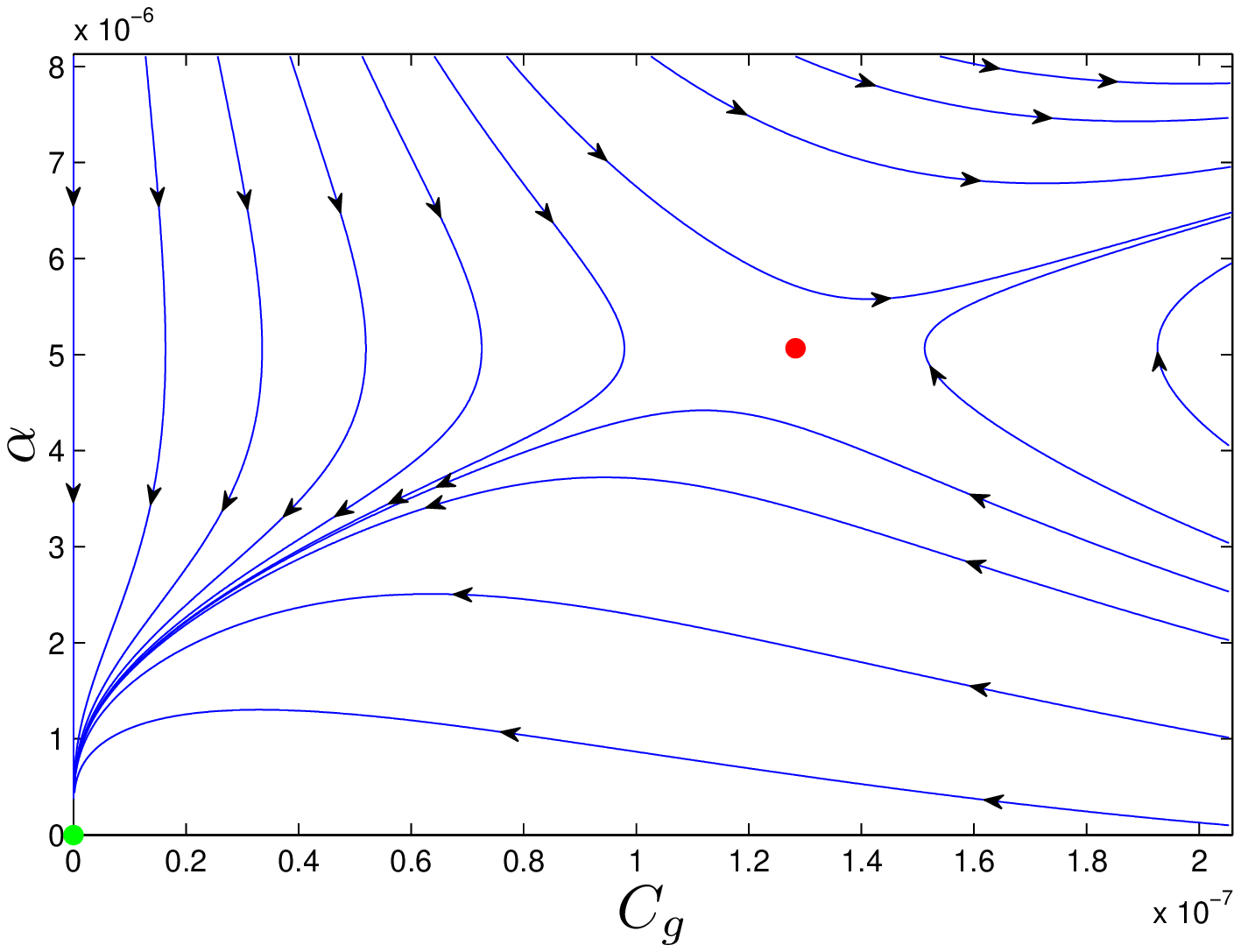}\label{WeakCouplingFlowGPx}}
\hspace{7.0ex} \subfigure{
\includegraphics[width=2.6in]{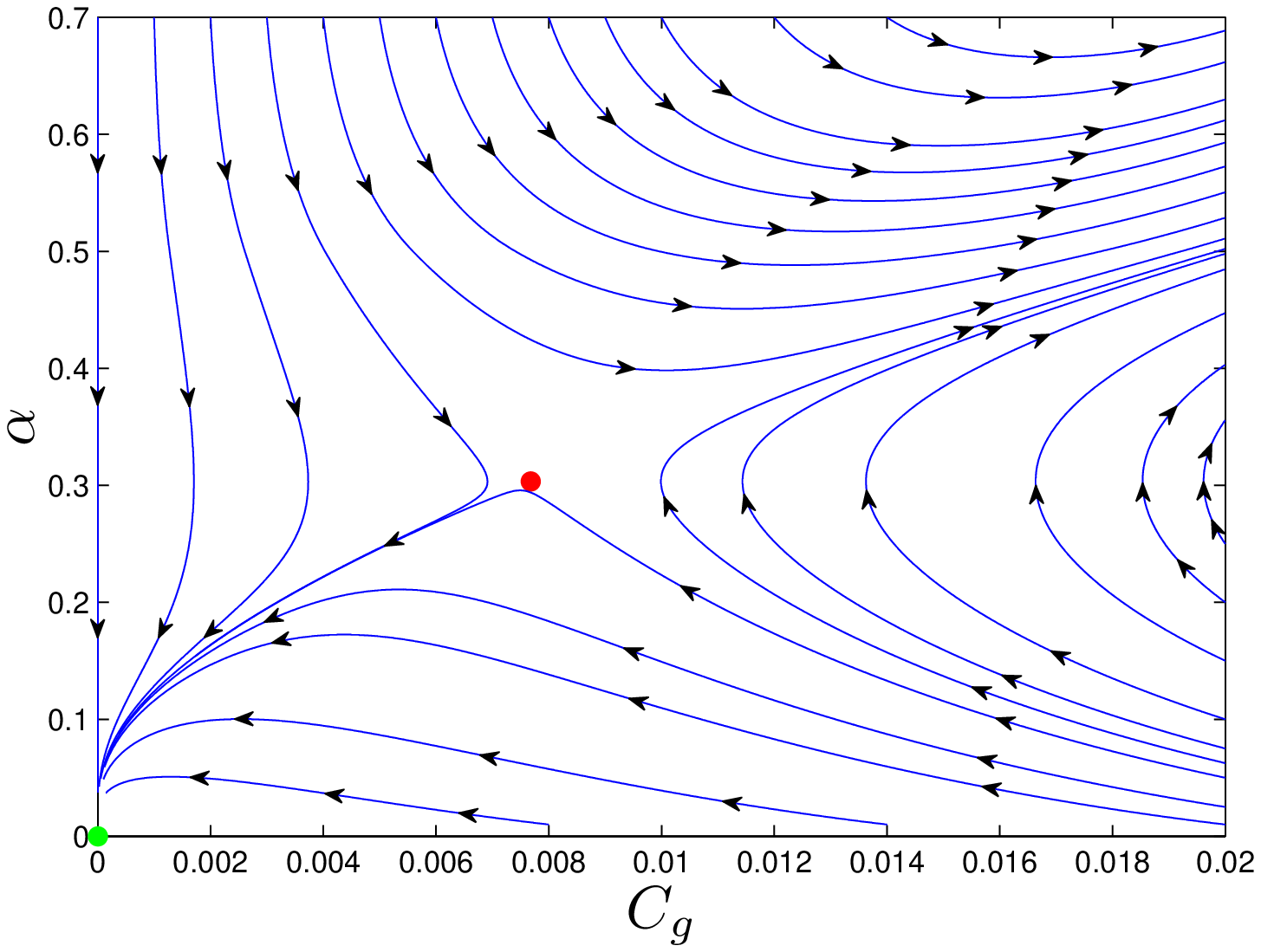}\label{WeakCouplingFlowGPy}}\vspace{-0.1cm}
\caption{Flow diagram in the space spanned by $\alpha$ and $C_{g}$
for $\tau_x$-component of random gauge potential is shown in (a) and
$\tau_y$-component in (b), where $\alpha$ in the weak coupling
limit. The red and purple points represent the Gaussian fixed point
and unstable finite fixed point, respectively. Here, the fermion
flavor is assumed to be $N = 2$.}
\end{figure*}

\subsection{Random gauge potential}\label{Sec_Ran_chem}

In the case of random gauge potential, the flow equation for $C_g$
is
\begin{eqnarray}
\frac{dC_{g}}{dl} = (2\gamma_{D} - \gamma_A - \gamma_v)C_g
\end{eqnarray}
in the strong coupling limit. Here, $\gamma_{D}$ take different
values for the $\tau_x$- and $\tau_y$-components. However, since
$\gamma_{D} > (\gamma_A + \gamma_v)/2$ for both components, it is
clear that $C_g(l)$ increases monotonously as $l$ grows, and tends
to diverge in the limit $l \rightarrow +\infty$. Therefore, although
random gauge potential is marginal in the non-interacting limit, it
becomes relevant due to strong Coulomb interaction. Moreover, the
parameters $A$ and $v$ are driven to vanish at the lowest energy.
From Eq.~(\ref{Eq.RG_alpha}), we see that $\alpha_N$ also increases
indefinitely as $l\longrightarrow +\infty$. Thus the effective
strength of Coulomb interaction is remarkably enhanced by random
gauge potential, and indeed becomes a relevant perturbation. As
already explained in the above discussions, one reasonable
interpretation of the indefinite increase of $\alpha_N$ and $C_{g}$
with growing $l$ is that the system becomes insulating.

We finally consider the weak coupling limit. For the
$\tau_x$-component of random gauge potential, the flow diagram is
depicted in Fig.~\ref{WeakCouplingFlowGPx}, and that of
$\tau_y$-component in Fig.~\ref{WeakCouplingFlowGPy}. Both of these
two flow diagrams contain a stable Gaussian fixed point and an
unstable finite fixed point. The RG flows of $\alpha_N$ and $C_g$
are analogous to those obtained in the cases of random chemical
potential and random, which has been discussed in great detail in
the last two subsections.

\section{Summary and discussion}\label{Sec_summary}

We have studied the quantum phase transitions and non-FL behaviors
induced by the interplay of Coulomb interaction and disorder in a 2D
SDSM in which the fermion dispersion is linear in one direction and
quadratic in the other. After performing extensive RG calculations,
we have showed that Coulomb interaction and disorder can
substantially affect each other, which then leads to a series of
interesting phases and transitions between distinct phases. The
concrete results obtained in the presence of three types of
disorders have already been summarized at the end of
Sev.~\ref{Sec_intro}, and thus are not repeated here.

Due to technical difficulties, we are not able to handle the case in
which the Coulomb interaction strength $\alpha_N$ is neither small
nor large. The results obtained in this work are applicable only to
the weak and strong coupling limits of Coulomb interaction.

The RG analysis have showed that the parameter $\alpha_N$ for
Coulomb interaction can flow to very large values due to its
interplay with disorders. A reasonable explanation for this behavior
is that certain quantum phase transition happens, which induces an
instability of the system and prevents $\alpha_N$ from really
flowing to infinity. However, the nature of such a phase transition
is still not clear. Generically, the 2D SDSM may become an excitonic
insulator \cite{Khve01, Khve04, Gorbar02, Liu09, Hands08, Drut09,
Khve09, Gamayun10, Kotov12, Wang12, Fischer13}, or even a Wigner
lattice \cite{Ceperley1989}. Further studies are certainly needed to
identify the true ground state of such an extremely correlated
fermion system. To examine whether it is possible to form an
excitonic insulator, one should go beyond the perturbative expansion
method, and study the excitonic insulating transition by using the
self-consistent Dyson-Schwinger equation approach \cite{Khve01,
Khve04, Gorbar02, Liu09, Khve09, Gamayun10, Kotov12, Wang12,
Fischer13}. When the Coulomb interaction is sufficiently strong, the
fermions may acquire a dynamically generated mass term, $\propto m
\tau_z$, which then drives the system to enter into an excitonic
insulating phase \cite{Khve01, Khve04, Gorbar02, Liu09, Hands08,
Drut09, Khve09, Gamayun10, Kotov12, Wang12, Fischer13}. This work is
now in progress, and will be presented in a separate paper
\cite{Wang16}. Large-scale numerical simulations are also expected
to provide useful information \cite{Hands08, Drut09}.

\section*{ACKNOWLEDGEMENTS}

The authors acknowledge the financial support by the National
Natural Science Foundation of China under Grants 11274286, 11574285,
11504379, and 11375168.

\end{document}